\documentclass[pra, twocolumn, english, superscriptaddress,floatfix, nofootinbib, citeautoscript, nobibnotes, longbibliography]{revtex4-2}

\usepackage{import}
\usepackage[utf8]{inputenc}
\usepackage[T1]{fontenc}
\usepackage{cmap}
\usepackage{graphicx}
\usepackage{mathtools}

\usepackage{setspace}

\usepackage[hyperfootnotes=false,breaklinks=true]{hyperref}
\hypersetup{
 bookmarksopen=true,
 bookmarksopenlevel=1,
 colorlinks=true,
 linkcolor=blue,
 anchorcolor=blue,
 citecolor=blue,
 filecolor=blue,
 urlcolor=blue,
 pdfpagemode=UseOutlines,
 pdfstartview={XYZ null null 1},
 linktocpage=true,
}

\usepackage{amsmath}
\allowdisplaybreaks  
\usepackage{amssymb}
\usepackage{amstext}
\usepackage{amsfonts}
\usepackage{mathrsfs}
\usepackage{amsthm}
\usepackage{dsfont}
\usepackage{bm}
\usepackage{braket}
\usepackage{physics}
\usepackage{enumitem}
\usepackage[
 capitalise,
]{cleveref}

\usepackage{pgfplots}
\pgfplotsset{compat=newest}
\usepackage[english]{babel}
\usepackage{algcompatible}
\newcounter{algorithm}
\renewcommand{\thealgorithm}{\arabic{algorithm}}

\crefname{algorithm}{Alg.}{Algs.}
\Crefname{algorithm}{Algorithm}{Algorithms}

\theoremstyle{plain}

\theoremstyle{definition}

\newcommand{\mycomment}[1]{}

\usepackage{relsize}
\usepackage{cancel}
\makeatletter
\def\blfootnote{\xdef\@thefnmark{}\@footnotetext}
\makeatother

\begin{document}

\blfootnote{This manuscript has been authored by UT-Battelle, LLC, under Contract No. DE-AC0500OR22725 with the U.S. Department of Energy. The United States Government retains and the publisher, by accepting the article for publication, acknowledges that the United States Government retains a non-exclusive, paid-up, irrevocable, worldwide license to publish or reproduce the published form of this manuscript, or allow others to do so, for the United States Government purposes. The Department of Energy will provide public access to these results of federally sponsored research in accordance with the DOE Public Access Plan.}

\title{Hamiltonian Learning at Scale}

\author{Nathan Johnson}  
\email{njjohnson0406@gmail.com}

\affiliation{Washington University in Saint Louis, Physics Department, Saint Louis, MO}
\affiliation{Computational Sciences and Engineering Division, 
Oak Ridge National Laboratory, 
Oak Ridge, Tennessee 37831, USA}

\author{Eugene Dumitrescu}  
\email{dumitrescuef@ornl.gov}
\affiliation{Computational Sciences and Engineering Division, 
Oak Ridge National Laboratory, 
Oak Ridge, Tennessee 37831, USA}

\date{\today}
\begin{abstract}
Learning a quantum system's Hamiltonian is crucial for understanding and controlling its dynamics and has recently become a topic of widespread interest. To understand the learning protocol's error tolerances, i.e. its stability in the presence of inevitable errors, this work utilizes tensor network techniques to emulate Hamiltonian learning workflows at scale and with noise. Specifically, we employed a Hamiltonian learning technique based on approximate stationary states which are constructed using matrix product tools. We provide analytic bounds on the Hamiltonian learning estimation errors and perform numerical simulations that highlight learning error's stability under two families of errors. Using our workflow, we are able to scale up the protocol and learn mixed-field Ising model Hamiltonians of an $N=300$ site spin chain. By simulating the protocol at large scale, and empirically studying its practical limitations, our analysis of how errors affect the Hamiltonian learning process provides valuable lessons for future experiments. We conclude by discussing the avenues our work opens as well as future work that can support Hamiltonian learning experiments. 

\end{abstract}
\maketitle

\section{Introduction}
\label{sec:motivation}

Determining the microscopic interactions that determine quantum dynamics is an important task. This is especially true given the proliferation of quantum devices across the physical sciences. For example, these quantum devices are integral to efforts in next-generation sensing~\cite{Degen2017}, communication~\cite{Azuma}, and computation~\cite{Georgescu}. Regardless of the particular application, a data-based and device-level characterization of the physical components, including their intra-system quantum mechanical interactions, is crucial to characterization and engineered co-design. 

To tackle this important task, many different approaches to Hamiltonian learning have been developed. These approaches range in terms of their underlying methodologies and assumptions. For example, Hamiltonian learning has been discussed in terms of practical experimental errors~\cite{Wiebe2014, brahmachari2026}, as a route towards analog quantum simulation~\cite{Kraft2025}, Bayesian approaches~\cite{Wiebe2015, Evans2019}, neural network approaches~\cite{Castelano2024}, quantum dynamics~\cite{Zhang2014, Hou2017, Li2020,Wilde2022,Arunachalam2024, Sohail2026, Bluhm2026, zhou2026, singh2026}, quantum error correction \cite{Romanov2026}, experimental demonstrations~\cite{Hangleiter2024}, Heisenberg-limited scaling \cite{Hu2025,Baran2025}, fermions \cite{Ni2024}, bosons \cite{Li2024boson}, Gibbs states \cite{Gu2024,Artymowicz2024b,Haah2024,Chen2025,Garcia-Pintos2024,bakshi2026}, and field theories \cite{Ott2024}. Learning methods have also been extended to Lindbladian learning \cite{Dumi2020, Olsacher2025, Heightman2026,arad2026}, which generalizes the protocol to additionally learn interactions that couple the quantum system of interest to their outside environments.

As the number of physical qubits increases in quantum devices, practical issues such as error analysis become increasingly important. For example, some protocols that were amenable at small system sizes~\cite{Hangleiter2024}, or even tomography itself~\cite{Kokail2021}, become difficult to experimentally implement at scale due to their unfavorable learning complexity. This has led to more approximate learning techniques~\cite{Guo2025}. By simulating Hamiltonian learning at scale, this work we tackles the important issue of studying the error scaling itself and its impact on the quality of Hamiltonian learning solutions. In doing so, our motivation is to provide practical guidance for future experiments regarding the feasibility of specific learning protocols. 

Our work is organized as follows. In Sec.~\ref{sec:methods} we outline our methodology. This includes the particular Hamiltonian learning protocol of choice, as well as our tensor network methods to simulate the protocol at scale. Next Sec.~\ref{sec:results} describes our main results. These include numerical simulations of the protocol on up to 300 qubits and the error scaling analysis. Finally, in Sec.~\ref{sec:conclusions}, we conclude by discussing open directions and the future research that will be required to realize successful Hamiltonian learning experiments at scale. 

\section{Methods}
\label{sec:methods}
\subsection{Hamiltonian Learning Protocol}
\label{sec:protocol}

As we have discussed in the introduction, there are a variety of complimentary methods for learning a quantum Hamiltonian. One set of techniques involves the use of certain constraints, like symmetries, and relies on locality for computational efficiency. For example, there is one set of approaches that can recover Hamiltonians from steady states, of the relevant quantum dynamics, by using stationary or conserved quantities to construct and solve a homogeneous correlation matrix~\cite{Bairey2019, Ranard2019, Bairey2020, Dumi2020}. Thus, our protocol begins by assuming access to a thermal (Gibbs) state $\rho(\beta,H) =\frac{ e^{-\beta H}}{\rm{Tr}[e^{-\beta H}]}$. The protocol therefore applies to experimental systems at finite temperature. In the next section we will describe a scalable approach to write modeling such states in the presence of errors. 

To learn the Hamiltonian, where $H=\sum_jc_jS_j$ is assumed to be a linear combination of local Pauli operators and $\vec{c}$ is the vector of coefficients in a particular basis, we then construct a matrix of commutators based on the Heisenberg equations of motion $\langle\dot{O}\rangle=-\frac{i}{\hbar}\langle[O,H]\rangle=-\sum_{j}\frac{c_ji}{\hbar}\langle[O,S_j]\rangle$. Since, for a stationary state, $\langle\dot{O}\rangle=0$ we can write down a system of equations in matrix form $A\vec{c}=\vec{0}$, where $A_{ij}=\langle[O_i,S_j]\rangle$ and $\{O_i\}$ is the observable basis. For local Hamiltonians, this basis will also be constructed from the collection of $j$-local Pauli operators.

The task of Hamiltonian learning then reduces to finding a vector $\vec{c}$, up to a scalar which can be fit from additional known experimental data, such that $A\vec{c}=\vec{0}$. The formal solution to the homogeneous linear equation then arises from linear algebra as the task of finding a null vector corresponding to a zero eigen- or singular-value of the matrix $A$. That is, given the matrix $A$, we can decompose it via the singular value decomposition (SVD) $A=U\Sigma V^{\dagger}$ with the row of $V^\dagger$ associated with the smallest singular value being the estimated solution.

In a realistic experiment, various types of errors will occur. These could include, for example, state preparation and measurement errors, drift in the temperature of the refrigerating apparatus, time dependent interactions between the qubits and their environment. In general, all these errors will contribute and their source may be unknown. To simulate unknown, average case errors that inevitably occur in experiments, we would therefore like to consider the learning protocol in terms of generic, unstructured errors. We will therefore model errors in two types of ways. First, as described in Sec.~\ref{sec:MPO}, we will use density operator approximation errors as a proxy for generic, unknown noise. Later, in Sec.~\ref{sec:results}, we will additionally consider stochastic errors on the correlation matrix which are independent of MPO truncation errors. We will return to the particular details of the Hamiltonian chosen and the learning results in Sec.~\ref{sec:results}.

\subsection{Matrix Product Construction}
\label{sec:MPO}

To scale our numerical computations, Alg.~\ref{alg:MPOconstruction} describes how we construct approximate thermal, at a temperature $T=\beta^{-1}$, states using matrix product operator (MPO) representations. Our approximate construction utilizes a Taylor series approximation, MPO compression, and renormalization. For small $\beta$, the Taylor series of the Gibbs operator \( G(\beta,H) = e^{-\beta H} = \sum_j \frac{(-\beta H)^j}{j!} \) converges rapidly. We first build $H$ as an MPO using the ITensors library ~\cite{SciPostPhysCodeb.4-r0.3, SciPostPhysCodeb.4}. In practice, for local Hamiltonians, $H$ can be written as an MPO with a small bond dimension $\chi$ and this fact underpins our construction. 

Following Alg.~\ref{alg:MPOconstruction}, higher powers of $H$ are constructed by taking the product of MPOs and then compressing them with a cutoff in the singular values. That is, we sweep across the chain contracting the physical indices in the MPO-MPO product which results in a new MPO. We then, again, sweep across the chain discarding all singular values smaller than a truncation-error cutoff $\varepsilon$ which adaptively reduces the bond dimension of the resulting product. As per the Taylor series expansion, the MPO corresponding to the $k^{th}$ power of $H$ is multiplied by a scalar coefficient $-\beta/k$. These MPOs are iteratively summed (as a direct sum) and the summands are again truncated with a cutoff $\varepsilon$ after each additive step. Finally, the trace is computed and the state is renormalized such that $\text{Tr}[\rho]=1$ is a valid density operator. Physical observables are then computed with respect to the resulting matrix product operator as $\langle \hat{O} \rangle = {\rm Tr}[ \hat{O} \rho ]$.

\begin{figure}[t]
\refstepcounter{algorithm}
\label{alg:MPOconstruction}

\hrule
\vspace{3pt}

\noindent\textbf{Algorithm \thealgorithm.\ Density Operator Synthesis}

\vspace{3pt}
\hrule
\vspace{4pt}

\begin{algorithmic}[1]
    \STATE $\mathds{1} =$ MPO(ComplexF64, sites, "Id")
    \STATE $T =$ copy($\mathds{1}$)
    \STATE $\rho =$ copy($\mathds{1}$)
    \FOR{$k = 1:K$}
        \STATE $T$ = apply($T$, $-\beta H/k$)
        \STATE truncate!($T$; cutoff $= \varepsilon$)
        \STATE $\rho \bigoplus = T$
        \STATE truncate!($\rho$; cutoff $= \varepsilon$)
    \ENDFOR
    \STATE $Z = \text{Tr}[\rho]$
    \STATE $\rho *= 1/Z$
\end{algorithmic}

\vspace{4pt}
\hrule
\end{figure}

We briefly describe the approximation errors in greater detail. Consider a Hamiltonian $H$ acting on $N$ sites spanned by a basis $\ket{\bm{s}}$, 
\begin{equation}
    H = \sum_{\bm{s, s'}} H_{\bm{s, s'}} \ket{\bm{s}}\bra{\bm{s'}}.
\end{equation}
The MPO representation of $H$ exposes its correlation structure across a bi-partition. That is, split the physical indices into the components to the left and right of a bipartition $\bm{s} = \{ \bm{s_L}, \bm{s_R}\}$. Truncation is taken, via intermediary auxiliary bonds, with respect to the Schmidt decomposition of the matrix $H_{(\bm{s_L},\bm{s_L}'),(\bm{s_R},\bm{s_R'})}$. That is, taking orthonormal operator bases on the left ($\{ L_\alpha \}$) and right ($\{ R_\gamma \}$), we have $H = \sum c_{\alpha, \gamma}  L_\alpha \otimes R_\gamma$. Truncating the $H$ MPO means taking the SVD of $c$ and subsequently discarding the singular values, below a threshold $\varepsilon$, which results in $\tilde{c}$ as the orthonormal core of $\tilde{H}$~\cite{SCHOLLWOCK201196}. 

The approximation error is in the Frobenius norm as we now detail. Using \[||H-\tilde{H}||_F = \sqrt{\text{Tr}[(H -  \tilde{H})^\dagger (H -  \tilde{H})]}\] we see that this error may be computed as the trace of matrix product operators as $||H-\tilde{H}||_F^2 = \text{Tr}[H^\dagger H] + \text{Tr}[\tilde{H}^\dagger \tilde{H} ] - 2 \text{Tr}[\tilde{H}^\dagger H]$ and the relative error is $\frac{||H-\tilde{H}||_F}{||H||_F}$. This error is then propagated into the construction of $e^{-\beta H}$ in terms of polynomial expansion in $H$. 

Thus, taken altogether, our density operator synthesis error is comprised of i) truncating the Taylor expansion, ii) truncation via MPO compression, and iii) renormalization. Finally, we are interested in the error of observables, evaluated with respect to our approximate density operator, as they comprise the correlation matrix $A$. While the protocol will work for a range of temperatures, with our choice of $\beta = 0.7$ the Taylor expansion error $\epsilon_{\rm{K}}=||G(\beta,H) - G_K(\beta,H)|| = || \sum_{k=K+1}^\infty (-\beta H)^k/k! ||$ is modest due to rapid Taylor series convergence. The second error is the compression error $\epsilon_C$. Although this error comes from the steps 5-8 in Alg.~\ref{alg:MPOconstruction}, we can summarize it as $\epsilon_C = ||G_K(\beta,H) - \tilde{G}_K(\beta,H)||$. The third source of error is renormalization. That is, the ideal normalization constant is $Z = \text{Tr}[G]$ while the normalization constant we utilize is $\tilde{Z}_K = \text{Tr}[\tilde{G}_K]$. Since trace is a linear function, these constants are simply the sum over the trace at each order. In App.~\ref{sec:trace_errors} we provide a table highlighting the trace errors for an $N=10$ model with a variety of cutoffs. 

\section{Results}
\label{sec:results}

To test Hamiltonian learning at scale we select a Ising model with transverse and longitudinal fields, 
\begin{equation}
    \label{eq:H}
    H=J\sum_{i=1}^{N-1}\sigma^z_i\sigma^z_{i+1}+\sum_{i=1}^N(h_z\sigma^z_i+h_x\sigma^x_i),
\end{equation}
where we have set $J=1$, $h_z=0.5$, and $h_x=0.7$. We then followed the learning protocol, outlined in Sec.~\ref{sec:protocol}. We take the right vector associated with the lowest singular value, encoding $\vec{c}_{\text{sim}}$, and then computed the relative error $\Delta c = \frac{||\vec{c}_{\text{sim}}-\vec{c}_{\text{true}}||}{||\vec{c}_{\text{true}}||}$ of the learned values with respect to the true interactions encoded by Eq.~\ref{eq:H}. 

\begin{figure}
    \centering
        \includegraphics[width=\linewidth]{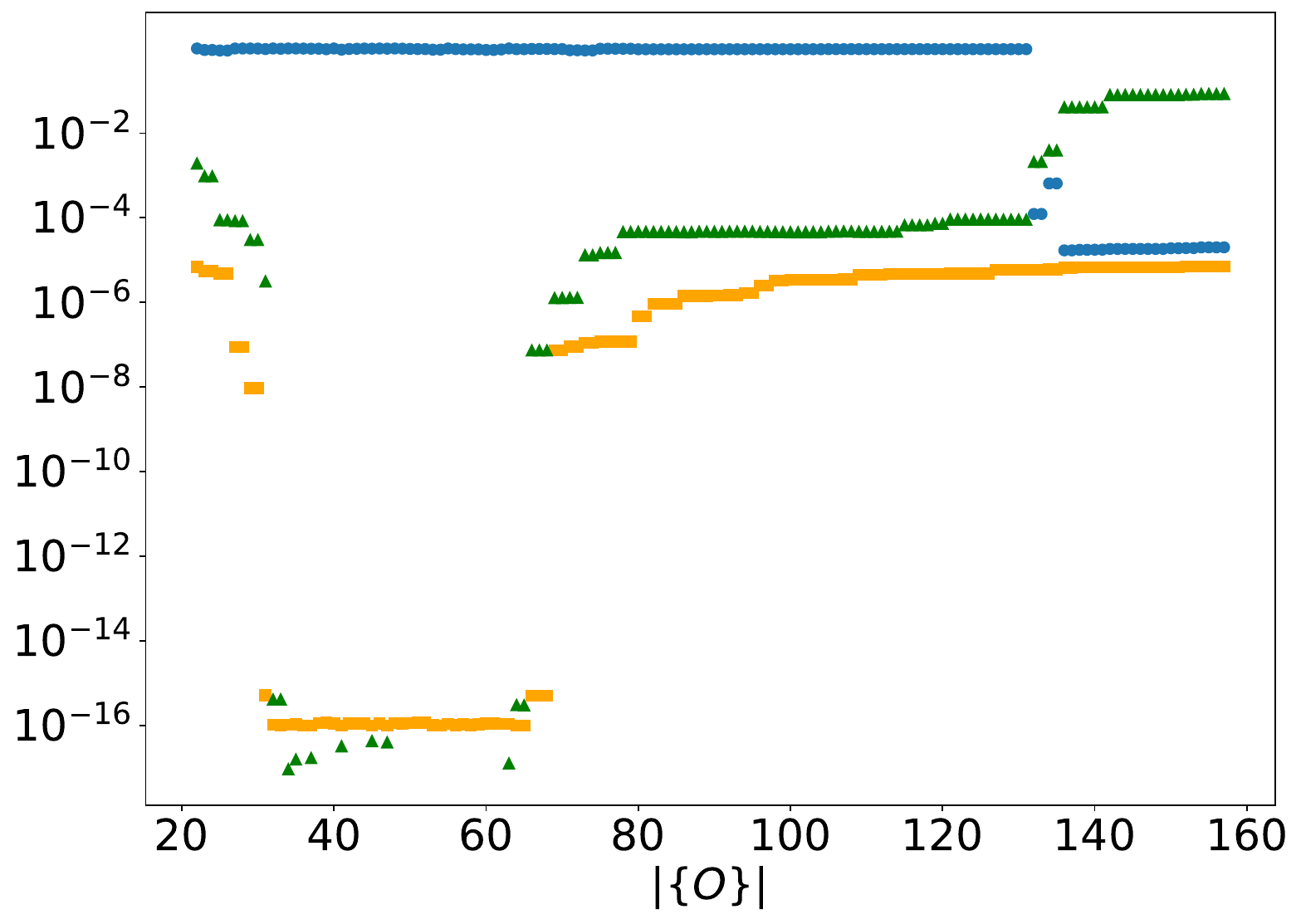}
        \caption{Hamiltonian learning data as a function of increasing operator basis cardinality $|\{ O\}|$ for $N=5$. The error in the Hamiltonian learning solution $\Delta c$ (blue circles) falls to $\sim 1\times10^{-5}$ when the the spectral gap $\Delta_s$, between the lowest and next singular values, (green triangles) is well defined, i.e., when $\Delta_s\gg s_0$ is much larger than the lowest singular value $s_0$ (orange squares).
        }        
        \label{fig:Singvals}
\end{figure}

Figure~\ref{fig:Singvals} illustrates the learning protocol and the intuition behind our forthcoming error analysis. In this figure the lowest singular value is denoted by the orange squares which are generally a small number and it is not clear whether they should be taken as a correct zero singular value. 

This question, of whether to accept a candidate solution, is important in the presence of noise. The green triangles indicate the gap between the lowest singular value and the next largest singular value. The blue squares indicate the relative learning error $\Delta c$. Note that, as we increased the operator basis for the learning procedure, we see that the singular values develops a gap $\Delta_s\gg s_0$ which indicates the recovery of the correct solution.

Let us now examine the error's impact within our the Hamiltonian learning protocol. Start with the correlation matrix with elements $A_{ij}=\text{Tr}(\rho[O_i,S_j])$. In practice, due to errors, one will measure a correlation matrix $\tilde{A}$ with a singular value distribution that differs from the ideal correlation matrix and its singular values. 

One source of error, described in Sec.~\ref{sec:MPO}, comes from error in the MPO approximation of the density matrix $E=\rho-\rho_{\text{ideal}}$. Since the trace is a linear operator, the element-wise errors, $\Delta A_{ij}=\text{Tr}[\rho[O_i,S_j]]-\text{Tr}[\tilde{\rho}[O_i,S_j]]=\text{Tr}[E[O_i,S_j]]$, are directly proportional to the MPO approximation error and the commutator $[O_i,S_j]$. 

Let us now derive upper bounds on the relative solution error. The measured matrix elements are all coefficients of the expansion of the density operator. Likewise, the error can be expanded in terms of Pauli operators $E=\sum_i c_i \hat{P_i}$. Each coefficient $c_i$ appears as a number $f_i(j,k)$ that just depends on the localities $j$ of the local observables and $k$ of the terms in the Hamiltonian. This is because the local Pauli operators with non-vanishing commutators, to give a Pauli operator $\hat{P_i}$, share support an odd number of times that scales at most with the Hamiltonian and selected operator $O_i$'s locality. Denote the maximum of all $f_i$ as $f$ and note that $||A_E||_F\leq f(j,k)||E||_F$.  

Once the error in the correlation matrix $A$ is bounded, one needs to apply perturbation theory to bound how the null vector and learned Hamiltonian $\vec{c}$ changes. First, note that $\vec{0}\approx A\vec{c}=(A_{\text{ideal}}+A_E)(\vec{c}_{\text{true}}+\Delta \vec{c})$ which to first order is $A_{\text{ideal}}\Delta \vec{c}+A_E\vec{c}_{\text{true}}$. One can either then apply a pseudo-inverse of $A_{\text{ideal}}$, or in the case where the nullspace of this matrix is one-dimensional, apply perturbation theory~\cite{Davis1970, Wedin1972} to get $\sin{\theta}\leq\frac{||A_E||}{\sigma_{n-1}}$ where $\theta$ is the angle between the true solution and the learned Hamiltonian coefficients and $\sigma_{n-1}$ is the corresponding smallest non-zero singular value of $A_{\text{ideal}}$. For small $\theta$, this implies $||\Delta \vec{c}||\leq\frac{||A_E||}{\sigma_{n-1}}\leq\frac{f(j,k)||E||_F}{\sigma_{n-1}}$ where $f(j,k)$ is only a function of locality and does not scale with the system size. 

\begin{figure}
        \centering
        \includegraphics[width=0.9\linewidth]{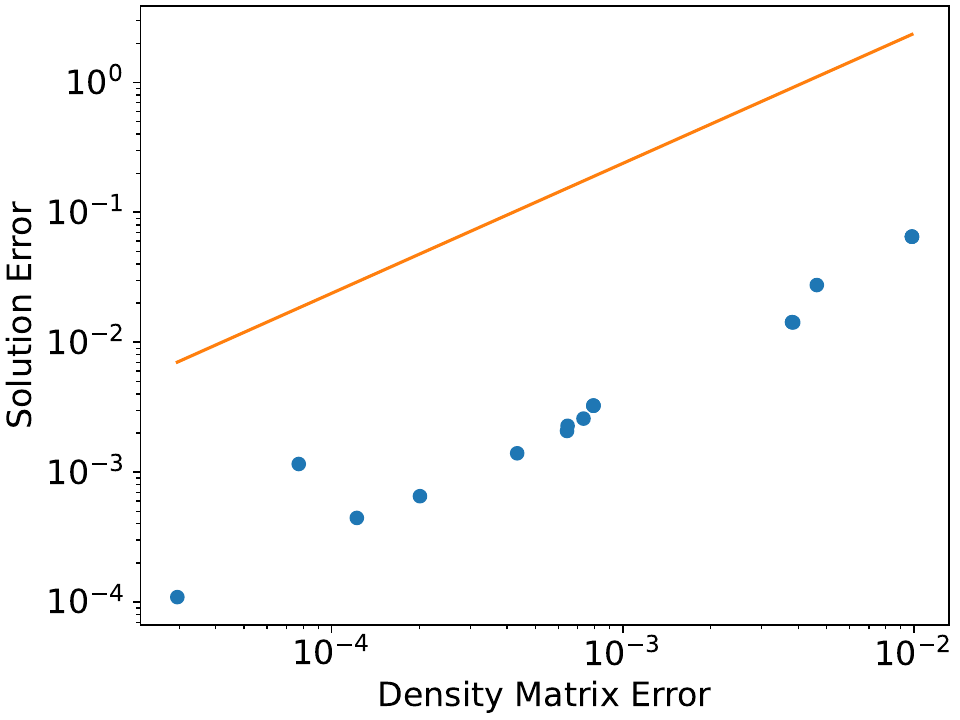}
        \caption{Scaling of solution error with error in $\rho$ at N=5. The error remains below the analytical upper bound.}
        \label{fig:Rho_err}
\end{figure}

To see how our bounds work in practice, we then plotted the numerically computed solution error versus the error in the initial thermal state for $N=5$. Note that we are only able to perform this analysis for small system sizes where explicit matrix exponential can be taken. Figure.~ \ref{fig:Rho_err} shows, on a log scale, that $\Delta c$ grows linearly with $||E||_F$ indicating a linear relationship between the variables. The errors all fall below the analytical bounds derived above. 

\begin{figure}
    \centering
        \includegraphics[width=\linewidth]{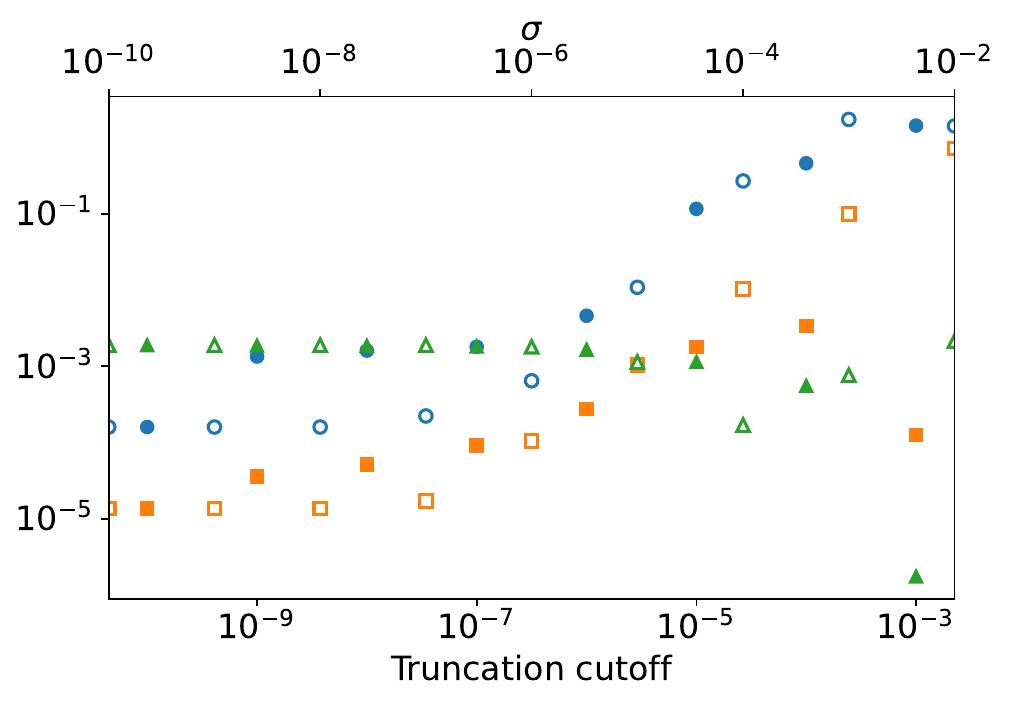}    
        \caption{Scaling of solution error $\Delta{c}$ (blue circles), the smallest singular value $s_0$ (orange squares), and the spectral gap $\Delta_s$ (green triangles) with the truncation cutoff for the $\rho$ MPO and the standard deviation $\sigma$ of Gaussian noise added to the correlation matrix at N=300. The trend with $\sigma$ is given by open shapes and the other trend by closed shapes. By $10^{-5}$ truncation error or $\sigma$, there is no longer a unique solution.}     
        \label{fig:Trunc_err}
\end{figure}  

As we scale the number of qubits, we are not able to calculate the exact error in the density matrix due to constraints on memory. No classical supercomputer can explicitly store all elements of a $2^{60}\times2^{60}$ density matrix. However, using Alg.~\ref{alg:MPOconstruction}'s MPO density operator construction, we now are able to scale the simulations of Hamiltonian learning up to $300$ qubit simulations. To understand the impact of noise on these simulations we perform two calculations. That is, we vary the MPO truncation $\varepsilon$ and also add Gaussian-distributed noise with standard deviation $\sigma$ as we scale to larger system size. These two complementary techniques allow us to continue studying the Hamiltonian learning errors in a regime where exact diagonalization is not feasible. 

The results of increasing truncation cutoff (solid symbols) and adding Gaussian noise to the correlation matrix (open symbols) are shown in Fig.~\ref{fig:Trunc_err}. We interpret the results as follows. The learning protocol is stable, as discussed above, under perturbations in the correlation matrix which do not mix the null solution vector with non-zero singular values. This is true in the parameter regimes where the noise is smaller than the spectral gap (green triangles). As per Weyl's theorem~\cite{Weyl1912, Stewart1990}, we observe that the magnitude of the lowest singular value increases linearly with both sources of noise (orange squares). Note that, at low $\sigma$, the errors do not change until $10^{-8}$, where they are no longer negligible compared to the smallest MPO truncation errors that we allow, and start following about a linear trend. 

By $10^{-5}$ truncation error, or $\sigma=10^{-5}$, the learned Hamiltonian is no longer accurate. The minimum singular values (orange squares) grow more quickly with respect to $\sigma$ (open symbols) than with truncation error (solid symbols). With the truncation error $\varepsilon$, there is what appears to be a piecewise, it is constant between $10^{-9}$ and $10^{-7}$, linear trend for the solution error within $\varepsilon\in[10^{-10},10^{-4}]$. The exact reason for this is not established, but the overall trend is that the solution error increase with respect to the truncation cutoff is slower than with $\sigma$. This could be because the types of errors we are introducing in the two different methods are quite different. Truncation cutoff throws away low weight correlations in the density matrix, which results in more structured errors, whereas Gaussian-distributed noise is fundamentally random with i.i.d. statistics that follow the law of large numbers for the $300$ qubit correlation matrix. In either case, $\sigma=10^{-5}$ or $\varepsilon=10^{-5}$ is the threshold where the learning protocol breaks. These results constrain the magnitudes of errors which are permitted, in experiments, in order for the Hamiltonian learning protocol to yield accurate estimates. While our results show that small errors are required, it is interesting to see that the size of the errors, for $300$ qubit systems is similar to earlier estimates for Hamiltonian learning at  scales which are two orders of magnitude smaller than considered in this work~\cite{Dumi2020}. 

\section{Conclusions}
\label{sec:conclusions}
Motivated by the growing scale of quantum hardware devices, this work scaled numerical Hamiltonian learning simulations to $300$ qubits. The Hamiltonian learning protocol we utilized is a standard protocol that makes use of a quantum system's conserved quantities of motion. To realize these simulations, we have utilized tensor network techniques based on the theory of matrix product operators and used them to build thermal steady states in terms of a Taylor expansion in the Hamiltonian. 

Our simulations shed light on a few complementary aspects of Hamiltonian learning. Firstly, with tensor network approximation errors in synthesizing the thermal density operator, the protocol was successful with relative Hamiltonian learning reconstruction errors around $10^{-4}$ for 300 qubits. Secondly, the simulations allowed us to study the stability of the learning protocol and its sensitivity to errors. Since modern quantum devices are plagued by errors (decoherence, dissipation, readout errors, and more) this study at scale is critical to assessing the viability of learning protocols in practice. Our results empirically justify previous work's assertions that the learning protocol is stable when errors are small compared to the gap between the lowest singular values of a factorized correlation matrix. This work paves the path for new research in a variety of directions that involve both classical and quantum computation as well as open quantum systems. 

Additional complementary ways to characterize the multi-parameter Hamiltonian learning errors in the constructed solution, akin to the approach in Ref.~\citenum{Garcia-Pintos2024}, can come from information-theoretical analysis based on the quantum Fisher information. This would give a rigorous differential geometry perspective for how error may propagate from different system parameters, like the field strength in our model. Then, one can use the Cramer-Rao bound, $\text{Var}(\theta)\geq \frac{1}{I_F(\theta)}$, to bound the variance in those parameters which are the coefficients of the Hamiltonian. 

Regarding open quantum systems, our Hamiltonian learning calculations at scale should be extended to Lindbladian learning~\cite{Heightman2026,arad2026}. This would generalize our error analysis to the critical case where quantum systems are coupled to Markovian reservoirs. In the realm of quantum computers, if successful, this protocol could serve as a certificate of a lack of non-Markovian dynamics. This is expected to become increasingly true for error corrected systems and could provide empirical evidence of this conjecture. Lastly, the classical simulation of quantum systems coupled to reservoirs with DMRG-methods has recently gained traction~\cite{Chan2005, Casagrande_2021, Yamamoto2022, Zhong25}. However, due to the formulation of the problem, this is more difficult than the closed-systems DMRG and often lacks certificates. Hence, such protocols could benefit from a verification stage. 

Lastly, our work and methods could be extended in a variety of directions. For example, our calculations could be extended to systems in two-dimensions, to fermionic systems, and to Floquet systems in (quasi-) equilibrium. For each of these systems, distinct computational tools that are appropriate to that setting should be utilized~\cite{VanDamme2024, Dai2025, vanthilt2026, Mickiewicz2026}. Overall, tensor network's ability to scale calculations has proved valuable across the physical sciences. Future research should make use of this resource in order to improve and validate Hamiltonian learning a wide variety of quantum many-body systems. 

\section{Acknowledgments}

E. D. is supported by the U.S. Department of Energy, Office of Science, Advanced Scientific Computing Research, Early Career Award under contract number ERKJ420. N. J. was supported by the U.S. Department of Energy, Office of Science, Office of Workforce Development for Teachers and Scientists (WDTS) under the Science Undergraduate Laboratory Internships (SULI) program hosted at the Oak Ridge National Laboratory and administered by the Oak Ridge Institute for Science and Education.

\appendix

\section{Trace Errors}
In this section we present the effects of various cutoffs on the trace of powers of the Hamiltonian operator. The thermal state is a polynomial, up to a truncation cutoff $K$, in these powers. 
\label{sec:trace_errors}
\begin{table*}[t]
\centering
\begin{tabular}{c|ccccc}
\hline
$\ell$ &
Dense $\mathrm{Tr}[H^\ell]$ &
Cutoff $0$ &
Cutoff $10^{-10}$ &
Cutoff $10^{-8}$ &
Cutoff $10^{-6}$ \\
\hline
0  & $1.02400000\times 10^{3}$  & $1.02400000\times 10^{3}$  & $1.02400000\times 10^{3}$  & $1.02400000\times 10^{3}$  & $1.02400000\times 10^{3}$ \\
1  & $2.13162821\times 10^{-14}$ & $-1.98908681\times 10^{-13}$ & $-1.98908681\times 10^{-13}$ & $-1.98908681\times 10^{-13}$ & $-1.98908681\times 10^{-13}$ \\
2  & $1.36955121\times 10^{3}$  & $1.36955121\times 10^{3}$  & $1.36955121\times 10^{3}$  & $1.36955121\times 10^{3}$  & $1.36955121\times 10^{3}$ \\
3  & $1.71944493\times 10^{1}$  & $1.71944493\times 10^{1}$  & $1.71944493\times 10^{1}$  & $1.71944493\times 10^{1}$  & $1.71942730\times 10^{1}$ \\
4  & $4.91442031\times 10^{3}$  & $4.91442031\times 10^{3}$  & $4.91442031\times 10^{3}$  & $4.91442031\times 10^{3}$  & $4.91442283\times 10^{3}$ \\
5  & $2.92049651\times 10^{2}$  & $2.92049651\times 10^{2}$  & $2.92049651\times 10^{2}$  & $2.92049796\times 10^{2}$  & $2.92167600\times 10^{2}$ \\
6  & $2.64354914\times 10^{4}$  & $2.64354914\times 10^{4}$  & $2.64354914\times 10^{4}$  & $2.64354990\times 10^{4}$  & $2.64346330\times 10^{4}$ \\
7  & $4.52359268\times 10^{3}$  & $4.52359268\times 10^{3}$  & $4.52359266\times 10^{3}$  & $4.52338555\times 10^{3}$  & $4.53652306\times 10^{3}$ \\
8  & $1.80387889\times 10^{5}$  & $1.80387889\times 10^{5}$  & $1.80387866\times 10^{5}$  & $1.80391087\times 10^{5}$  & $1.80081469\times 10^{5}$ \\
9  & $6.52025227\times 10^{4}$  & $6.52025227\times 10^{4}$  & $6.52023394\times 10^{4}$  & $6.52007906\times 10^{4}$  & $6.53205726\times 10^{4}$ \\
10 & $1.44587642\times 10^{6}$  & $1.44587642\times 10^{6}$  & $1.44587690\times 10^{6}$  & $1.44584505\times 10^{6}$  & $1.44582852\times 10^{6}$ \\
\hline
\end{tabular}
\caption{Trace moments $\mathrm{Tr}[H^\ell]$ for the $N=10$ random-field Ising Hamiltonian. All imaginary parts are zero to printed precision.}
\label{tab:ising-trace-moments}
\end{table*}


\begin{thebibliography}{55}%
\makeatletter
\providecommand \@ifxundefined [1]{%
 \@ifx{#1\undefined}
}%
\providecommand \@ifnum [1]{%
 \ifnum #1\expandafter \@firstoftwo
 \else \expandafter \@secondoftwo
 \fi
}%
\providecommand \@ifx [1]{%
 \ifx #1\expandafter \@firstoftwo
 \else \expandafter \@secondoftwo
 \fi
}%
\providecommand \natexlab [1]{#1}%
\providecommand \enquote  [1]{``#1''}%
\providecommand \bibnamefont  [1]{#1}%
\providecommand \bibfnamefont [1]{#1}%
\providecommand \citenamefont [1]{#1}%
\providecommand \href@noop [0]{\@secondoftwo}%
\providecommand \href [0]{\begingroup \@sanitize@url \@href}%
\providecommand \@href[1]{\@@startlink{#1}\@@href}%
\providecommand \@@href[1]{\endgroup#1\@@endlink}%
\providecommand \@sanitize@url [0]{\catcode `\\12\catcode `\$12\catcode
  `\&12\catcode `\#12\catcode `\^12\catcode `\_12\catcode `\%12\relax}%
\providecommand \@@startlink[1]{}%
\providecommand \@@endlink[0]{}%
\providecommand \url  [0]{\begingroup\@sanitize@url \@url }%
\providecommand \@url [1]{\endgroup\@href {#1}{\urlprefix }}%
\providecommand \urlprefix  [0]{URL }%
\providecommand \Eprint [0]{\href }%
\providecommand \doibase [0]{https://doi.org/}%
\providecommand \selectlanguage [0]{\@gobble}%
\providecommand \bibinfo  [0]{\@secondoftwo}%
\providecommand \bibfield  [0]{\@secondoftwo}%
\providecommand \translation [1]{[#1]}%
\providecommand \BibitemOpen [0]{}%
\providecommand \bibitemStop [0]{}%
\providecommand \bibitemNoStop [0]{.\EOS\space}%
\providecommand \EOS [0]{\spacefactor3000\relax}%
\providecommand \BibitemShut  [1]{\csname bibitem#1\endcsname}%
\let\auto@bib@innerbib\@empty
\bibitem [{\citenamefont {Degen}\ \emph {et~al.}(2017)\citenamefont {Degen},
  \citenamefont {Reinhard},\ and\ \citenamefont {Cappellaro}}]{Degen2017}%
  \BibitemOpen
  \bibfield  {author} {\bibinfo {author} {\bibfnamefont {C.~L.}\ \bibnamefont
  {Degen}}, \bibinfo {author} {\bibfnamefont {F.}~\bibnamefont {Reinhard}},\
  and\ \bibinfo {author} {\bibfnamefont {P.}~\bibnamefont {Cappellaro}},\
  }\bibfield  {title} {\bibinfo {title} {Quantum sensing},\ }\href
  {https://doi.org/10.1103/RevModPhys.89.035002} {\bibfield  {journal}
  {\bibinfo  {journal} {Rev. Mod. Phys.}\ }\textbf {\bibinfo {volume} {89}},\
  \bibinfo {pages} {035002} (\bibinfo {year} {2017})}\BibitemShut {NoStop}%
\bibitem [{\citenamefont {Azuma}\ \emph {et~al.}(2023)\citenamefont {Azuma},
  \citenamefont {Economou}, \citenamefont {Elkouss}, \citenamefont {Hilaire},
  \citenamefont {Jiang}, \citenamefont {Lo},\ and\ \citenamefont
  {Tzitrin}}]{Azuma}%
  \BibitemOpen
  \bibfield  {author} {\bibinfo {author} {\bibfnamefont {K.}~\bibnamefont
  {Azuma}}, \bibinfo {author} {\bibfnamefont {S.~E.}\ \bibnamefont {Economou}},
  \bibinfo {author} {\bibfnamefont {D.}~\bibnamefont {Elkouss}}, \bibinfo
  {author} {\bibfnamefont {P.}~\bibnamefont {Hilaire}}, \bibinfo {author}
  {\bibfnamefont {L.}~\bibnamefont {Jiang}}, \bibinfo {author} {\bibfnamefont
  {H.-K.}\ \bibnamefont {Lo}},\ and\ \bibinfo {author} {\bibfnamefont
  {I.}~\bibnamefont {Tzitrin}},\ }\bibfield  {title} {\bibinfo {title} {Quantum
  repeaters: From quantum networks to the quantum internet},\ }\href
  {https://doi.org/10.1103/RevModPhys.95.045006} {\bibfield  {journal}
  {\bibinfo  {journal} {Rev. Mod. Phys.}\ }\textbf {\bibinfo {volume} {95}},\
  \bibinfo {pages} {045006} (\bibinfo {year} {2023})}\BibitemShut {NoStop}%
\bibitem [{\citenamefont {Georgescu}\ \emph {et~al.}(2014)\citenamefont
  {Georgescu}, \citenamefont {Ashhab},\ and\ \citenamefont {Nori}}]{Georgescu}%
  \BibitemOpen
  \bibfield  {author} {\bibinfo {author} {\bibfnamefont {I.~M.}\ \bibnamefont
  {Georgescu}}, \bibinfo {author} {\bibfnamefont {S.}~\bibnamefont {Ashhab}},\
  and\ \bibinfo {author} {\bibfnamefont {F.}~\bibnamefont {Nori}},\ }\bibfield
  {title} {\bibinfo {title} {Quantum simulation},\ }\href
  {https://doi.org/10.1103/RevModPhys.86.153} {\bibfield  {journal} {\bibinfo
  {journal} {Rev. Mod. Phys.}\ }\textbf {\bibinfo {volume} {86}},\ \bibinfo
  {pages} {153} (\bibinfo {year} {2014})}\BibitemShut {NoStop}%
\bibitem [{\citenamefont {Wiebe}\ \emph {et~al.}(2014)\citenamefont {Wiebe},
  \citenamefont {Granade}, \citenamefont {Ferrie},\ and\ \citenamefont
  {Cory}}]{Wiebe2014}%
  \BibitemOpen
  \bibfield  {author} {\bibinfo {author} {\bibfnamefont {N.}~\bibnamefont
  {Wiebe}}, \bibinfo {author} {\bibfnamefont {C.}~\bibnamefont {Granade}},
  \bibinfo {author} {\bibfnamefont {C.}~\bibnamefont {Ferrie}},\ and\ \bibinfo
  {author} {\bibfnamefont {D.~G.}\ \bibnamefont {Cory}},\ }\bibfield  {title}
  {\bibinfo {title} {Quantum hamiltonian learning using imperfect quantum
  resources},\ }\href {https://doi.org/10.1103/PhysRevA.89.042314} {\bibfield
  {journal} {\bibinfo  {journal} {Phys. Rev. A}\ }\textbf {\bibinfo {volume}
  {89}},\ \bibinfo {pages} {042314} (\bibinfo {year} {2014})},\ \Eprint
  {https://arxiv.org/abs/1311.5269} {arXiv:1311.5269} \BibitemShut {NoStop}%
\bibitem [{\citenamefont {Brahmachari}\ \emph {et~al.}(2026)\citenamefont
  {Brahmachari}, \citenamefont {Zhu}, \citenamefont {Marvian},\ and\
  \citenamefont {Tong}}]{brahmachari2026}%
  \BibitemOpen
  \bibfield  {author} {\bibinfo {author} {\bibfnamefont {S.}~\bibnamefont
  {Brahmachari}}, \bibinfo {author} {\bibfnamefont {S.}~\bibnamefont {Zhu}},
  \bibinfo {author} {\bibfnamefont {I.}~\bibnamefont {Marvian}},\ and\ \bibinfo
  {author} {\bibfnamefont {Y.}~\bibnamefont {Tong}},\ }\href
  {https://arxiv.org/abs/2601.10380} {\bibinfo {title} {Learning hamiltonians
  in the heisenberg limit with static single-qubit fields}} (\bibinfo {year}
  {2026}),\ \Eprint {https://arxiv.org/abs/2601.10380} {arXiv:2601.10380
  [quant-ph]} \BibitemShut {NoStop}%
\bibitem [{\citenamefont {Kraft}\ \emph {et~al.}(2025)\citenamefont {Kraft},
  \citenamefont {Joshi}, \citenamefont {Lam}, \citenamefont {Olsacher},
  \citenamefont {Kranzl}, \citenamefont {Franke}, \citenamefont {Joshi},
  \citenamefont {Blatt}, \citenamefont {Smerzi}, \citenamefont
  {Stilck~Fran{\c{c}}a}, \citenamefont {Vermersch}, \citenamefont {Kraus},
  \citenamefont {Roos},\ and\ \citenamefont {Zoller}}]{Kraft2025}%
  \BibitemOpen
  \bibfield  {author} {\bibinfo {author} {\bibfnamefont {T.}~\bibnamefont
  {Kraft}}, \bibinfo {author} {\bibfnamefont {M.~K.}\ \bibnamefont {Joshi}},
  \bibinfo {author} {\bibfnamefont {W.}~\bibnamefont {Lam}}, \bibinfo {author}
  {\bibfnamefont {T.}~\bibnamefont {Olsacher}}, \bibinfo {author}
  {\bibfnamefont {F.}~\bibnamefont {Kranzl}}, \bibinfo {author} {\bibfnamefont
  {J.}~\bibnamefont {Franke}}, \bibinfo {author} {\bibfnamefont {L.~K.}\
  \bibnamefont {Joshi}}, \bibinfo {author} {\bibfnamefont {R.}~\bibnamefont
  {Blatt}}, \bibinfo {author} {\bibfnamefont {A.}~\bibnamefont {Smerzi}},
  \bibinfo {author} {\bibfnamefont {D.}~\bibnamefont {Stilck~Fran{\c{c}}a}},
  \bibinfo {author} {\bibfnamefont {B.}~\bibnamefont {Vermersch}}, \bibinfo
  {author} {\bibfnamefont {B.}~\bibnamefont {Kraus}}, \bibinfo {author}
  {\bibfnamefont {C.~F.}\ \bibnamefont {Roos}},\ and\ \bibinfo {author}
  {\bibfnamefont {P.}~\bibnamefont {Zoller}},\ }\href@noop {} {\bibinfo {title}
  {Bounded-error quantum simulation via hamiltonian and lindbladian learning}}
  (\bibinfo {year} {2025}),\ \Eprint {https://arxiv.org/abs/2511.23392}
  {arXiv:2511.23392} \BibitemShut {NoStop}%
\bibitem [{\citenamefont {Wiebe}\ \emph {et~al.}(2015)\citenamefont {Wiebe},
  \citenamefont {Granade},\ and\ \citenamefont {Cory}}]{Wiebe2015}%
  \BibitemOpen
  \bibfield  {author} {\bibinfo {author} {\bibfnamefont {N.}~\bibnamefont
  {Wiebe}}, \bibinfo {author} {\bibfnamefont {C.}~\bibnamefont {Granade}},\
  and\ \bibinfo {author} {\bibfnamefont {D.~G.}\ \bibnamefont {Cory}},\
  }\bibfield  {title} {\bibinfo {title} {Quantum bootstrapping via compressed
  quantum hamiltonian learning},\ }\href
  {https://doi.org/10.1088/1367-2630/17/2/022005} {\bibfield  {journal}
  {\bibinfo  {journal} {New J. Phys.}\ }\textbf {\bibinfo {volume} {17}},\
  \bibinfo {pages} {022005} (\bibinfo {year} {2015})},\ \Eprint
  {https://arxiv.org/abs/1409.1524} {arXiv:1409.1524} \BibitemShut {NoStop}%
\bibitem [{\citenamefont {Evans}\ \emph {et~al.}(2019)\citenamefont {Evans},
  \citenamefont {Harper},\ and\ \citenamefont {Flammia}}]{Evans2019}%
  \BibitemOpen
  \bibfield  {author} {\bibinfo {author} {\bibfnamefont {T.~J.}\ \bibnamefont
  {Evans}}, \bibinfo {author} {\bibfnamefont {R.}~\bibnamefont {Harper}},\ and\
  \bibinfo {author} {\bibfnamefont {S.~T.}\ \bibnamefont {Flammia}},\
  }\href@noop {} {\bibinfo {title} {Scalable bayesian hamiltonian learning}}
  (\bibinfo {year} {2019}),\ \Eprint {https://arxiv.org/abs/1912.07636}
  {arXiv:1912.07636} \BibitemShut {NoStop}%
\bibitem [{\citenamefont {Castelano}\ \emph {et~al.}(2024)\citenamefont
  {Castelano}, \citenamefont {Cunha}, \citenamefont {Luiz}, \citenamefont
  {de~Jesus~Napolitano}, \citenamefont {Prado},\ and\ \citenamefont
  {Fanchini}}]{Castelano2024}%
  \BibitemOpen
  \bibfield  {author} {\bibinfo {author} {\bibfnamefont {L.~K.}\ \bibnamefont
  {Castelano}}, \bibinfo {author} {\bibfnamefont {I.}~\bibnamefont {Cunha}},
  \bibinfo {author} {\bibfnamefont {F.~S.}\ \bibnamefont {Luiz}}, \bibinfo
  {author} {\bibfnamefont {R.}~\bibnamefont {de~Jesus~Napolitano}}, \bibinfo
  {author} {\bibfnamefont {M.~V. d.~S.}\ \bibnamefont {Prado}},\ and\ \bibinfo
  {author} {\bibfnamefont {F.~F.}\ \bibnamefont {Fanchini}},\ }\bibfield
  {title} {\bibinfo {title} {Combining physics-informed neural networks with
  the freezing mechanism for general hamiltonian learning},\ }\href
  {https://doi.org/10.1103/PhysRevA.110.032607} {\bibfield  {journal} {\bibinfo
   {journal} {Phys. Rev. A}\ }\textbf {\bibinfo {volume} {110}},\ \bibinfo
  {pages} {032607} (\bibinfo {year} {2024})}\BibitemShut {NoStop}%
\bibitem [{\citenamefont {Zhang}\ and\ \citenamefont
  {Sarovar}(2014)}]{Zhang2014}%
  \BibitemOpen
  \bibfield  {author} {\bibinfo {author} {\bibfnamefont {J.}~\bibnamefont
  {Zhang}}\ and\ \bibinfo {author} {\bibfnamefont {M.}~\bibnamefont
  {Sarovar}},\ }\bibfield  {title} {\bibinfo {title} {Quantum hamiltonian
  identification from measurement time traces},\ }\href
  {https://doi.org/10.1103/PhysRevLett.113.080401} {\bibfield  {journal}
  {\bibinfo  {journal} {Phys. Rev. Lett.}\ }\textbf {\bibinfo {volume} {113}},\
  \bibinfo {pages} {080401} (\bibinfo {year} {2014})},\ \Eprint
  {https://arxiv.org/abs/1401.5780} {arXiv:1401.5780} \BibitemShut {NoStop}%
\bibitem [{\citenamefont {Hou}\ \emph {et~al.}(2017)\citenamefont {Hou},
  \citenamefont {Li},\ and\ \citenamefont {Long}}]{Hou2017}%
  \BibitemOpen
  \bibfield  {author} {\bibinfo {author} {\bibfnamefont {S.-y.}\ \bibnamefont
  {Hou}}, \bibinfo {author} {\bibfnamefont {H.}~\bibnamefont {Li}},\ and\
  \bibinfo {author} {\bibfnamefont {G.-l.}\ \bibnamefont {Long}},\ }\bibfield
  {title} {\bibinfo {title} {Experimental quantum hamiltonian identification
  from measurement time traces},\ }\href
  {https://doi.org/10.1016/j.scib.2017.06.005} {\bibfield  {journal} {\bibinfo
  {journal} {Sci. Bull.}\ }\textbf {\bibinfo {volume} {62}},\ \bibinfo {pages}
  {863} (\bibinfo {year} {2017})},\ \Eprint {https://arxiv.org/abs/1410.3940}
  {arXiv:1410.3940} \BibitemShut {NoStop}%
\bibitem [{\citenamefont {Li}\ \emph {et~al.}(2020)\citenamefont {Li},
  \citenamefont {Zou},\ and\ \citenamefont {Hsieh}}]{Li2020}%
  \BibitemOpen
  \bibfield  {author} {\bibinfo {author} {\bibfnamefont {Z.}~\bibnamefont
  {Li}}, \bibinfo {author} {\bibfnamefont {L.}~\bibnamefont {Zou}},\ and\
  \bibinfo {author} {\bibfnamefont {T.~H.}\ \bibnamefont {Hsieh}},\ }\bibfield
  {title} {\bibinfo {title} {Hamiltonian tomography via quantum quench},\
  }\href {https://doi.org/10.1103/PhysRevLett.124.160502} {\bibfield  {journal}
  {\bibinfo  {journal} {Phys. Rev. Lett.}\ }\textbf {\bibinfo {volume} {124}},\
  \bibinfo {pages} {160502} (\bibinfo {year} {2020})},\ \Eprint
  {https://arxiv.org/abs/1912.09492} {arXiv:1912.09492} \BibitemShut {NoStop}%
\bibitem [{\citenamefont {Wilde}\ \emph {et~al.}(2026)\citenamefont {Wilde},
  \citenamefont {Kshetrimayum}, \citenamefont {Roth}, \citenamefont
  {Hangleiter}, \citenamefont {Sweke},\ and\ \citenamefont
  {Eisert}}]{Wilde2022}%
  \BibitemOpen
  \bibfield  {author} {\bibinfo {author} {\bibfnamefont {F.}~\bibnamefont
  {Wilde}}, \bibinfo {author} {\bibfnamefont {A.}~\bibnamefont {Kshetrimayum}},
  \bibinfo {author} {\bibfnamefont {I.}~\bibnamefont {Roth}}, \bibinfo {author}
  {\bibfnamefont {D.}~\bibnamefont {Hangleiter}}, \bibinfo {author}
  {\bibfnamefont {R.}~\bibnamefont {Sweke}},\ and\ \bibinfo {author}
  {\bibfnamefont {J.}~\bibnamefont {Eisert}},\ }\bibfield  {title} {\bibinfo
  {title} {Scalably learning one-dimensional quantum many-body hamiltonians
  from dynamical data},\ }\href {https://doi.org/10.1088/2058-9565/ae6fe3}
  {\bibfield  {journal} {\bibinfo  {journal} {Quantum Science and Technology}\
  }\textbf {\bibinfo {volume} {11}},\ \bibinfo {pages} {035002} (\bibinfo
  {year} {2026})}\BibitemShut {NoStop}%
\bibitem [{\citenamefont {Arunachalam}\ \emph {et~al.}(2024)\citenamefont
  {Arunachalam}, \citenamefont {Dutt},\ and\ \citenamefont
  {Escudero~Guti{\'e}rrez}}]{Arunachalam2024}%
  \BibitemOpen
  \bibfield  {author} {\bibinfo {author} {\bibfnamefont {S.}~\bibnamefont
  {Arunachalam}}, \bibinfo {author} {\bibfnamefont {A.}~\bibnamefont {Dutt}},\
  and\ \bibinfo {author} {\bibfnamefont {F.}~\bibnamefont
  {Escudero~Guti{\'e}rrez}},\ }\href@noop {} {\bibinfo {title} {Testing and
  learning structured quantum hamiltonians}} (\bibinfo {year} {2024}),\ \Eprint
  {https://arxiv.org/abs/2411.00082} {arXiv:2411.00082} \BibitemShut {NoStop}%
\bibitem [{\citenamefont {Sohail}\ \emph {et~al.}(2026)\citenamefont {Sohail},
  \citenamefont {Sudharshan}, \citenamefont {Pradhan},\ and\ \citenamefont
  {Rao}}]{Sohail2026}%
  \BibitemOpen
  \bibfield  {author} {\bibinfo {author} {\bibfnamefont {M.~A.}\ \bibnamefont
  {Sohail}}, \bibinfo {author} {\bibfnamefont {R.~R.}\ \bibnamefont
  {Sudharshan}}, \bibinfo {author} {\bibfnamefont {S.~S.}\ \bibnamefont
  {Pradhan}},\ and\ \bibinfo {author} {\bibfnamefont {A.}~\bibnamefont {Rao}},\
  }\href@noop {} {\bibinfo {title} {Quantum hamiltonian learning using
  time-resolved measurement data and its application to gene regulatory network
  inference}} (\bibinfo {year} {2026}),\ \Eprint
  {https://arxiv.org/abs/2602.19496} {arXiv:2602.19496} \BibitemShut {NoStop}%
\bibitem [{\citenamefont {Bluhm}\ \emph {et~al.}(2026)\citenamefont {Bluhm},
  \citenamefont {Caro}, \citenamefont {Escudero~Guti{\'e}rrez}, \citenamefont
  {Lee}, \citenamefont {Oufkir}, \citenamefont {Rouz{\'e}},\ and\ \citenamefont
  {Shin}}]{Bluhm2026}%
  \BibitemOpen
  \bibfield  {author} {\bibinfo {author} {\bibfnamefont {A.}~\bibnamefont
  {Bluhm}}, \bibinfo {author} {\bibfnamefont {M.~C.}\ \bibnamefont {Caro}},
  \bibinfo {author} {\bibfnamefont {F.}~\bibnamefont {Escudero~Guti{\'e}rrez}},
  \bibinfo {author} {\bibfnamefont {J.}~\bibnamefont {Lee}}, \bibinfo {author}
  {\bibfnamefont {A.}~\bibnamefont {Oufkir}}, \bibinfo {author} {\bibfnamefont
  {C.}~\bibnamefont {Rouz{\'e}}},\ and\ \bibinfo {author} {\bibfnamefont
  {M.}~\bibnamefont {Shin}},\ }\href@noop {} {\bibinfo {title} {Certifying and
  learning local quantum hamiltonians}} (\bibinfo {year} {2026}),\ \Eprint
  {https://arxiv.org/abs/2603.29809} {arXiv:2603.29809} \BibitemShut {NoStop}%
\bibitem [{\citenamefont {Zhou}\ and\ \citenamefont {Gong}(2026)}]{zhou2026}%
  \BibitemOpen
  \bibfield  {author} {\bibinfo {author} {\bibfnamefont {T.}~\bibnamefont
  {Zhou}}\ and\ \bibinfo {author} {\bibfnamefont {W.}~\bibnamefont {Gong}},\
  }\href {https://arxiv.org/abs/2606.19486} {\bibinfo {title} {Optimal
  ansatz-free hamiltonian learning in situ}} (\bibinfo {year} {2026}),\ \Eprint
  {https://arxiv.org/abs/2606.19486} {arXiv:2606.19486 [quant-ph]} \BibitemShut
  {NoStop}%
\bibitem [{\citenamefont {Singh}\ \emph {et~al.}(2026)\citenamefont {Singh},
  \citenamefont {Barros},\ and\ \citenamefont {Li}}]{singh2026}%
  \BibitemOpen
  \bibfield  {author} {\bibinfo {author} {\bibfnamefont {N.}~\bibnamefont
  {Singh}}, \bibinfo {author} {\bibfnamefont {K.}~\bibnamefont {Barros}},\ and\
  \bibinfo {author} {\bibfnamefont {X.~S.}\ \bibnamefont {Li}},\ }\href
  {https://arxiv.org/abs/2605.12801} {\bibinfo {title} {Fast and stable
  gradient approximation for bilinear forms of hermitian matrix functions}}
  (\bibinfo {year} {2026}),\ \Eprint {https://arxiv.org/abs/2605.12801}
  {arXiv:2605.12801 [math.NA]} \BibitemShut {NoStop}%
\bibitem [{\citenamefont {Romanov}\ \emph {et~al.}(2026)\citenamefont
  {Romanov}, \citenamefont {Ivashkov}, \citenamefont {Gong}, \citenamefont
  {Kannan}, \citenamefont {Gu}, \citenamefont {Hu},\ and\ \citenamefont
  {Yelin}}]{Romanov2026}%
  \BibitemOpen
  \bibfield  {author} {\bibinfo {author} {\bibfnamefont {N.}~\bibnamefont
  {Romanov}}, \bibinfo {author} {\bibfnamefont {P.}~\bibnamefont {Ivashkov}},
  \bibinfo {author} {\bibfnamefont {W.}~\bibnamefont {Gong}}, \bibinfo {author}
  {\bibfnamefont {I.}~\bibnamefont {Kannan}}, \bibinfo {author} {\bibfnamefont
  {A.}~\bibnamefont {Gu}}, \bibinfo {author} {\bibfnamefont {H.-Y.}\
  \bibnamefont {Hu}},\ and\ \bibinfo {author} {\bibfnamefont {S.~F.}\
  \bibnamefont {Yelin}},\ }\href@noop {} {\bibinfo {title} {Learning arbitrary
  lindbladians with quantum error correction}} (\bibinfo {year} {2026}),\
  \Eprint {https://arxiv.org/abs/2606.18188} {arXiv:2606.18188 [quant-ph]}
  \BibitemShut {NoStop}%
\bibitem [{\citenamefont {Hangleiter}\ \emph {et~al.}(2024)\citenamefont
  {Hangleiter}, \citenamefont {Roth}, \citenamefont {Fuksa}, \citenamefont
  {Eisert},\ and\ \citenamefont {Roushan}}]{Hangleiter2024}%
  \BibitemOpen
  \bibfield  {author} {\bibinfo {author} {\bibfnamefont {D.}~\bibnamefont
  {Hangleiter}}, \bibinfo {author} {\bibfnamefont {I.}~\bibnamefont {Roth}},
  \bibinfo {author} {\bibfnamefont {J.}~\bibnamefont {Fuksa}}, \bibinfo
  {author} {\bibfnamefont {J.}~\bibnamefont {Eisert}},\ and\ \bibinfo {author}
  {\bibfnamefont {P.}~\bibnamefont {Roushan}},\ }\bibfield  {title} {\bibinfo
  {title} {Robustly learning the hamiltonian dynamics of a superconducting
  quantum processor},\ }\href {https://doi.org/10.1038/s41467-024-52629-3}
  {\bibfield  {journal} {\bibinfo  {journal} {Nature Communications}\ }\textbf
  {\bibinfo {volume} {15}},\ \bibinfo {pages} {9595} (\bibinfo {year}
  {2024})}\BibitemShut {NoStop}%
\bibitem [{\citenamefont {Hu}\ \emph {et~al.}(2025)\citenamefont {Hu},
  \citenamefont {Ma}, \citenamefont {Gong}, \citenamefont {Ye}, \citenamefont
  {Tong}, \citenamefont {Flammia},\ and\ \citenamefont {Yelin}}]{Hu2025}%
  \BibitemOpen
  \bibfield  {author} {\bibinfo {author} {\bibfnamefont {H.-Y.}\ \bibnamefont
  {Hu}}, \bibinfo {author} {\bibfnamefont {M.}~\bibnamefont {Ma}}, \bibinfo
  {author} {\bibfnamefont {W.}~\bibnamefont {Gong}}, \bibinfo {author}
  {\bibfnamefont {Q.}~\bibnamefont {Ye}}, \bibinfo {author} {\bibfnamefont
  {Y.}~\bibnamefont {Tong}}, \bibinfo {author} {\bibfnamefont {S.~T.}\
  \bibnamefont {Flammia}},\ and\ \bibinfo {author} {\bibfnamefont {S.~F.}\
  \bibnamefont {Yelin}},\ }\bibfield  {title} {\bibinfo {title} {Ansatz-free
  hamiltonian learning with heisenberg-limited scaling},\ }\href
  {https://doi.org/10.1103/PRXQuantum.6.040315} {\bibfield  {journal} {\bibinfo
   {journal} {PRX Quantum}\ }\textbf {\bibinfo {volume} {6}},\ \bibinfo {pages}
  {040315} (\bibinfo {year} {2025})},\ \Eprint
  {https://arxiv.org/abs/2502.11900} {arXiv:2502.11900} \BibitemShut {NoStop}%
\bibitem [{\citenamefont {Baran}\ and\ \citenamefont
  {Heightman}(2025)}]{Baran2025}%
  \BibitemOpen
  \bibfield  {author} {\bibinfo {author} {\bibfnamefont {B.}~\bibnamefont
  {Baran}}\ and\ \bibinfo {author} {\bibfnamefont {T.}~\bibnamefont
  {Heightman}},\ }\href@noop {} {\bibinfo {title} {Heisenberg-limited quantum
  hamiltonian learning via randomly spread product states}} (\bibinfo {year}
  {2025}),\ \Eprint {https://arxiv.org/abs/2507.21374} {arXiv:2507.21374}
  \BibitemShut {NoStop}%
\bibitem [{\citenamefont {Ni}\ \emph {et~al.}(2024)\citenamefont {Ni},
  \citenamefont {Li},\ and\ \citenamefont {Ying}}]{Ni2024}%
  \BibitemOpen
  \bibfield  {author} {\bibinfo {author} {\bibfnamefont {H.}~\bibnamefont
  {Ni}}, \bibinfo {author} {\bibfnamefont {H.}~\bibnamefont {Li}},\ and\
  \bibinfo {author} {\bibfnamefont {L.}~\bibnamefont {Ying}},\ }\href@noop {}
  {\bibinfo {title} {Quantum hamiltonian learning for the fermi-hubbard model}}
  (\bibinfo {year} {2024}),\ \Eprint {https://arxiv.org/abs/2312.17390}
  {arXiv:2312.17390} \BibitemShut {NoStop}%
\bibitem [{\citenamefont {Li}\ \emph {et~al.}(2024)\citenamefont {Li},
  \citenamefont {Tong}, \citenamefont {Ni}, \citenamefont {Gefen},\ and\
  \citenamefont {Ying}}]{Li2024boson}%
  \BibitemOpen
  \bibfield  {author} {\bibinfo {author} {\bibfnamefont {H.}~\bibnamefont
  {Li}}, \bibinfo {author} {\bibfnamefont {Y.}~\bibnamefont {Tong}}, \bibinfo
  {author} {\bibfnamefont {H.}~\bibnamefont {Ni}}, \bibinfo {author}
  {\bibfnamefont {T.}~\bibnamefont {Gefen}},\ and\ \bibinfo {author}
  {\bibfnamefont {L.}~\bibnamefont {Ying}},\ }\bibfield  {title} {\bibinfo
  {title} {Heisenberg-limited hamiltonian learning for interacting bosons},\
  }\href {https://doi.org/10.1038/s41534-024-00812-w} {\bibfield  {journal}
  {\bibinfo  {journal} {npj Quantum Inf.}\ }\textbf {\bibinfo {volume} {10}},\
  \bibinfo {pages} {83} (\bibinfo {year} {2024})},\ \Eprint
  {https://arxiv.org/abs/2307.04690} {arXiv:2307.04690} \BibitemShut {NoStop}%
\bibitem [{\citenamefont {Gu}\ \emph {et~al.}(2024)\citenamefont {Gu},
  \citenamefont {Cincio},\ and\ \citenamefont {Coles}}]{Gu2024}%
  \BibitemOpen
  \bibfield  {author} {\bibinfo {author} {\bibfnamefont {A.}~\bibnamefont
  {Gu}}, \bibinfo {author} {\bibfnamefont {{\L}.}~\bibnamefont {Cincio}},\ and\
  \bibinfo {author} {\bibfnamefont {P.~J.}\ \bibnamefont {Coles}},\ }\bibfield
  {title} {\bibinfo {title} {Practical hamiltonian learning with unitary
  dynamics and gibbs states},\ }\href
  {https://doi.org/10.1038/s41467-023-44008-1} {\bibfield  {journal} {\bibinfo
  {journal} {Nat. Commun.}\ }\textbf {\bibinfo {volume} {15}},\ \bibinfo
  {pages} {312} (\bibinfo {year} {2024})},\ \Eprint
  {https://arxiv.org/abs/2206.15464} {arXiv:2206.15464} \BibitemShut {NoStop}%
\bibitem [{\citenamefont {Artymowicz}(2024)}]{Artymowicz2024b}%
  \BibitemOpen
  \bibfield  {author} {\bibinfo {author} {\bibfnamefont {A.}~\bibnamefont
  {Artymowicz}},\ }\href@noop {} {\bibinfo {title} {Efficient hamiltonian
  learning from gibbs states}} (\bibinfo {year} {2024}),\ \Eprint
  {https://arxiv.org/abs/2403.18061} {arXiv:2403.18061} \BibitemShut {NoStop}%
\bibitem [{\citenamefont {Haah}\ \emph {et~al.}(2024)\citenamefont {Haah},
  \citenamefont {Kothari},\ and\ \citenamefont {Tang}}]{Haah2024}%
  \BibitemOpen
  \bibfield  {author} {\bibinfo {author} {\bibfnamefont {J.}~\bibnamefont
  {Haah}}, \bibinfo {author} {\bibfnamefont {R.}~\bibnamefont {Kothari}},\ and\
  \bibinfo {author} {\bibfnamefont {E.}~\bibnamefont {Tang}},\ }\bibfield
  {title} {\bibinfo {title} {Optimal learning of quantum hamiltonians from
  high-temperature gibbs states},\ }\href
  {https://doi.org/10.1038/s41567-023-02376-x} {\bibfield  {journal} {\bibinfo
  {journal} {Nat. Phys.}\ }\textbf {\bibinfo {volume} {20}},\ \bibinfo {pages}
  {1027} (\bibinfo {year} {2024})},\ \Eprint {https://arxiv.org/abs/2108.04842}
  {arXiv:2108.04842} \BibitemShut {NoStop}%
\bibitem [{\citenamefont {Chen}\ \emph {et~al.}(2025)\citenamefont {Chen},
  \citenamefont {Anshu},\ and\ \citenamefont {Nguyen}}]{Chen2025}%
  \BibitemOpen
  \bibfield  {author} {\bibinfo {author} {\bibfnamefont {C.-F.}\ \bibnamefont
  {Chen}}, \bibinfo {author} {\bibfnamefont {A.}~\bibnamefont {Anshu}},\ and\
  \bibinfo {author} {\bibfnamefont {Q.~T.}\ \bibnamefont {Nguyen}},\
  }\href@noop {} {\bibinfo {title} {Learning quantum gibbs states locally and
  efficiently}} (\bibinfo {year} {2025}),\ \Eprint
  {https://arxiv.org/abs/2504.02706} {arXiv:2504.02706} \BibitemShut {NoStop}%
\bibitem [{\citenamefont {Garc\'{\i}a-Pintos}\ \emph
  {et~al.}(2024)\citenamefont {Garc\'{\i}a-Pintos}, \citenamefont {Bharti},
  \citenamefont {Bringewatt}, \citenamefont {Dehghani}, \citenamefont
  {Ehrenberg}, \citenamefont {Yunger~Halpern},\ and\ \citenamefont
  {Gorshkov}}]{Garcia-Pintos2024}%
  \BibitemOpen
  \bibfield  {author} {\bibinfo {author} {\bibfnamefont {L.~P.}\ \bibnamefont
  {Garc\'{\i}a-Pintos}}, \bibinfo {author} {\bibfnamefont {K.}~\bibnamefont
  {Bharti}}, \bibinfo {author} {\bibfnamefont {J.}~\bibnamefont {Bringewatt}},
  \bibinfo {author} {\bibfnamefont {H.}~\bibnamefont {Dehghani}}, \bibinfo
  {author} {\bibfnamefont {A.}~\bibnamefont {Ehrenberg}}, \bibinfo {author}
  {\bibfnamefont {N.}~\bibnamefont {Yunger~Halpern}},\ and\ \bibinfo {author}
  {\bibfnamefont {A.~V.}\ \bibnamefont {Gorshkov}},\ }\bibfield  {title}
  {\bibinfo {title} {Estimation of hamiltonian parameters from thermal
  states},\ }\href {https://doi.org/10.1103/PhysRevLett.133.040802} {\bibfield
  {journal} {\bibinfo  {journal} {Phys. Rev. Lett.}\ }\textbf {\bibinfo
  {volume} {133}},\ \bibinfo {pages} {040802} (\bibinfo {year}
  {2024})}\BibitemShut {NoStop}%
\bibitem [{\citenamefont {Bakshi}\ \emph {et~al.}(2026)\citenamefont {Bakshi},
  \citenamefont {Liu}, \citenamefont {Moitra},\ and\ \citenamefont
  {Tang}}]{bakshi2026}%
  \BibitemOpen
  \bibfield  {author} {\bibinfo {author} {\bibfnamefont {A.}~\bibnamefont
  {Bakshi}}, \bibinfo {author} {\bibfnamefont {A.}~\bibnamefont {Liu}},
  \bibinfo {author} {\bibfnamefont {A.}~\bibnamefont {Moitra}},\ and\ \bibinfo
  {author} {\bibfnamefont {E.}~\bibnamefont {Tang}},\ }\href
  {https://arxiv.org/abs/2310.02243} {\bibinfo {title} {Learning quantum
  hamiltonians at any temperature in polynomial time}} (\bibinfo {year}
  {2026}),\ \Eprint {https://arxiv.org/abs/2310.02243} {arXiv:2310.02243
  [quant-ph]} \BibitemShut {NoStop}%
\bibitem [{\citenamefont {Ott}\ \emph {et~al.}(2024)\citenamefont {Ott},
  \citenamefont {Sirois}, \citenamefont {Sorey}, \citenamefont {Xu},
  \citenamefont {Chiesa} \emph {et~al.}}]{Ott2024}%
  \BibitemOpen
  \bibfield  {author} {\bibinfo {author} {\bibfnamefont {R.}~\bibnamefont
  {Ott}}, \bibinfo {author} {\bibfnamefont {A.-J.}\ \bibnamefont {Sirois}},
  \bibinfo {author} {\bibfnamefont {J.}~\bibnamefont {Sorey}}, \bibinfo
  {author} {\bibfnamefont {D.}~\bibnamefont {Xu}}, \bibinfo {author}
  {\bibfnamefont {A.}~\bibnamefont {Chiesa}}, \emph {et~al.},\ }\bibfield
  {title} {\bibinfo {title} {Hamiltonian learning in quantum field theories},\
  }\href {https://doi.org/10.1103/PhysRevResearch.6.043284} {\bibfield
  {journal} {\bibinfo  {journal} {Phys. Rev. Research}\ }\textbf {\bibinfo
  {volume} {6}},\ \bibinfo {pages} {043284} (\bibinfo {year} {2024})},\ \Eprint
  {https://arxiv.org/abs/2401.01308} {arXiv:2401.01308} \BibitemShut {NoStop}%
\bibitem [{\citenamefont {Dumitrescu}\ and\ \citenamefont
  {Lougovski}(2020)}]{Dumi2020}%
  \BibitemOpen
  \bibfield  {author} {\bibinfo {author} {\bibfnamefont {E.~F.}\ \bibnamefont
  {Dumitrescu}}\ and\ \bibinfo {author} {\bibfnamefont {P.}~\bibnamefont
  {Lougovski}},\ }\bibfield  {title} {\bibinfo {title} {Hamiltonian assignment
  for open quantum systems},\ }\href
  {https://doi.org/10.1103/PhysRevResearch.2.033251} {\bibfield  {journal}
  {\bibinfo  {journal} {Phys. Rev. Res.}\ }\textbf {\bibinfo {volume} {2}},\
  \bibinfo {pages} {033251} (\bibinfo {year} {2020})}\BibitemShut {NoStop}%
\bibitem [{\citenamefont {Olsacher}\ \emph {et~al.}(2025)\citenamefont
  {Olsacher}, \citenamefont {Kraft}, \citenamefont {Kokail}, \citenamefont
  {Kraus},\ and\ \citenamefont {Zoller}}]{Olsacher2025}%
  \BibitemOpen
  \bibfield  {author} {\bibinfo {author} {\bibfnamefont {T.}~\bibnamefont
  {Olsacher}}, \bibinfo {author} {\bibfnamefont {T.}~\bibnamefont {Kraft}},
  \bibinfo {author} {\bibfnamefont {C.}~\bibnamefont {Kokail}}, \bibinfo
  {author} {\bibfnamefont {B.}~\bibnamefont {Kraus}},\ and\ \bibinfo {author}
  {\bibfnamefont {P.}~\bibnamefont {Zoller}},\ }\bibfield  {title} {\bibinfo
  {title} {Hamiltonian and liouvillian learning in weakly-dissipative quantum
  many-body systems},\ }\href {https://doi.org/10.1088/2058-9565/ad9ed5}
  {\bibfield  {journal} {\bibinfo  {journal} {Quantum Science and Technology}\
  }\textbf {\bibinfo {volume} {10}},\ \bibinfo {pages} {015065} (\bibinfo
  {year} {2025})}\BibitemShut {NoStop}%
\bibitem [{\citenamefont {Heightman}\ \emph {et~al.}(2026)\citenamefont
  {Heightman}, \citenamefont {Aseguinolaza~Gallo}, \citenamefont {Jiang},
  \citenamefont {Saavedra}, \citenamefont {Ac{\'\i}n},\ and\ \citenamefont
  {P{\l}odzie{\'n}}}]{Heightman2026}%
  \BibitemOpen
  \bibfield  {author} {\bibinfo {author} {\bibfnamefont {T.}~\bibnamefont
  {Heightman}}, \bibinfo {author} {\bibfnamefont {R.}~\bibnamefont
  {Aseguinolaza~Gallo}}, \bibinfo {author} {\bibfnamefont {E.}~\bibnamefont
  {Jiang}}, \bibinfo {author} {\bibfnamefont {J.~R.~M.}\ \bibnamefont
  {Saavedra}}, \bibinfo {author} {\bibfnamefont {A.}~\bibnamefont
  {Ac{\'\i}n}},\ and\ \bibinfo {author} {\bibfnamefont {M.}~\bibnamefont
  {P{\l}odzie{\'n}}},\ }\href@noop {} {\bibinfo {title} {Lindbladian learning
  with neural differential equations}} (\bibinfo {year} {2026}),\ \Eprint
  {https://arxiv.org/abs/2603.07778} {arXiv:2603.07778} \BibitemShut {NoStop}%
\bibitem [{\citenamefont {Arad}\ \emph {et~al.}(2026)\citenamefont {Arad},
  \citenamefont {Chen}, \citenamefont {Guo}, \citenamefont {Rebentrost},\ and\
  \citenamefont {Yu}}]{arad2026}%
  \BibitemOpen
  \bibfield  {author} {\bibinfo {author} {\bibfnamefont {I.}~\bibnamefont
  {Arad}}, \bibinfo {author} {\bibfnamefont {Z.}~\bibnamefont {Chen}}, \bibinfo
  {author} {\bibfnamefont {N.}~\bibnamefont {Guo}}, \bibinfo {author}
  {\bibfnamefont {P.}~\bibnamefont {Rebentrost}},\ and\ \bibinfo {author}
  {\bibfnamefont {Z.}~\bibnamefont {Yu}},\ }\href
  {https://arxiv.org/abs/2606.20535} {\bibinfo {title} {Near-optimal learning
  of local lindbladians}} (\bibinfo {year} {2026}),\ \Eprint
  {https://arxiv.org/abs/2606.20535} {arXiv:2606.20535 [quant-ph]} \BibitemShut
  {NoStop}%
\bibitem [{\citenamefont {Kokail}\ \emph {et~al.}(2021)\citenamefont {Kokail},
  \citenamefont {{van Bijnen}}, \citenamefont {Elben}, \citenamefont
  {Vermersch},\ and\ \citenamefont {Zoller}}]{Kokail2021}%
  \BibitemOpen
  \bibfield  {author} {\bibinfo {author} {\bibfnamefont {C.}~\bibnamefont
  {Kokail}}, \bibinfo {author} {\bibfnamefont {R.}~\bibnamefont {{van
  Bijnen}}}, \bibinfo {author} {\bibfnamefont {A.}~\bibnamefont {Elben}},
  \bibinfo {author} {\bibfnamefont {B.}~\bibnamefont {Vermersch}},\ and\
  \bibinfo {author} {\bibfnamefont {P.}~\bibnamefont {Zoller}},\ }\bibfield
  {title} {\bibinfo {title} {Entanglement hamiltonian tomography in quantum
  simulation},\ }\href {https://doi.org/10.1038/s41567-021-01260-w} {\bibfield
  {journal} {\bibinfo  {journal} {Nat. Phys.}\ }\textbf {\bibinfo {volume}
  {17}},\ \bibinfo {pages} {936} (\bibinfo {year} {2021})},\ \Eprint
  {https://arxiv.org/abs/2009.09000} {arXiv:2009.09000} \BibitemShut {NoStop}%
\bibitem [{\citenamefont {Guo}\ \emph {et~al.}(2025)\citenamefont {Guo},
  \citenamefont {Wu}, \citenamefont {Ye}, \citenamefont {Zhang}, \citenamefont
  {Wang}, \citenamefont {Lian}, \citenamefont {Yao}, \citenamefont {Xu},
  \citenamefont {Zhang}, \citenamefont {Xu}, \citenamefont {Qi}, \citenamefont
  {Hou}, \citenamefont {He}, \citenamefont {Zhou},\ and\ \citenamefont
  {Duan}}]{Guo2025}%
  \BibitemOpen
  \bibfield  {author} {\bibinfo {author} {\bibfnamefont {S.-A.}\ \bibnamefont
  {Guo}}, \bibinfo {author} {\bibfnamefont {Y.-K.}\ \bibnamefont {Wu}},
  \bibinfo {author} {\bibfnamefont {J.}~\bibnamefont {Ye}}, \bibinfo {author}
  {\bibfnamefont {L.}~\bibnamefont {Zhang}}, \bibinfo {author} {\bibfnamefont
  {Y.}~\bibnamefont {Wang}}, \bibinfo {author} {\bibfnamefont {W.-Q.}\
  \bibnamefont {Lian}}, \bibinfo {author} {\bibfnamefont {R.}~\bibnamefont
  {Yao}}, \bibinfo {author} {\bibfnamefont {Y.-L.}\ \bibnamefont {Xu}},
  \bibinfo {author} {\bibfnamefont {C.}~\bibnamefont {Zhang}}, \bibinfo
  {author} {\bibfnamefont {Y.-Z.}\ \bibnamefont {Xu}}, \bibinfo {author}
  {\bibfnamefont {B.-X.}\ \bibnamefont {Qi}}, \bibinfo {author} {\bibfnamefont
  {P.-Y.}\ \bibnamefont {Hou}}, \bibinfo {author} {\bibfnamefont
  {L.}~\bibnamefont {He}}, \bibinfo {author} {\bibfnamefont {Z.-C.}\
  \bibnamefont {Zhou}},\ and\ \bibinfo {author} {\bibfnamefont {L.-M.}\
  \bibnamefont {Duan}},\ }\bibfield  {title} {\bibinfo {title} {Hamiltonian
  learning for 300 trapped ion qubits with long-range couplings},\ }\href
  {https://doi.org/10.1126/sciadv.adt4713} {\bibfield  {journal} {\bibinfo
  {journal} {Science Advances}\ }\textbf {\bibinfo {volume} {11}},\ \bibinfo
  {pages} {eadt4713} (\bibinfo {year} {2025})}\BibitemShut {NoStop}%
\bibitem [{\citenamefont {Bairey}\ \emph {et~al.}(2019)\citenamefont {Bairey},
  \citenamefont {Arad},\ and\ \citenamefont {Lindner}}]{Bairey2019}%
  \BibitemOpen
  \bibfield  {author} {\bibinfo {author} {\bibfnamefont {E.}~\bibnamefont
  {Bairey}}, \bibinfo {author} {\bibfnamefont {I.}~\bibnamefont {Arad}},\ and\
  \bibinfo {author} {\bibfnamefont {N.~H.}\ \bibnamefont {Lindner}},\
  }\bibfield  {title} {\bibinfo {title} {Learning a local hamiltonian from
  local measurements},\ }\href {https://doi.org/10.1103/PhysRevLett.122.020504}
  {\bibfield  {journal} {\bibinfo  {journal} {Physical Review Letters}\
  }\textbf {\bibinfo {volume} {122}},\ \bibinfo {pages} {020504} (\bibinfo
  {year} {2019})}\BibitemShut {NoStop}%
\bibitem [{\citenamefont {Ranard}\ and\ \citenamefont {Qi}(2019)}]{Ranard2019}%
  \BibitemOpen
  \bibfield  {author} {\bibinfo {author} {\bibfnamefont {D.}~\bibnamefont
  {Ranard}}\ and\ \bibinfo {author} {\bibfnamefont {X.-L.}\ \bibnamefont
  {Qi}},\ }\bibfield  {title} {\bibinfo {title} {Determining a local
  hamiltonian from a single eigenstate},\ }\href
  {https://doi.org/10.22331/q-2019-07-08-159} {\bibfield  {journal} {\bibinfo
  {journal} {Quantum}\ }\textbf {\bibinfo {volume} {3}},\ \bibinfo {pages}
  {159} (\bibinfo {year} {2019})}\BibitemShut {NoStop}%
\bibitem [{\citenamefont {Bairey}\ \emph {et~al.}(2020)\citenamefont {Bairey},
  \citenamefont {Guo}, \citenamefont {Poletti}, \citenamefont {Lindner},\ and\
  \citenamefont {Arad}}]{Bairey2020}%
  \BibitemOpen
  \bibfield  {author} {\bibinfo {author} {\bibfnamefont {E.}~\bibnamefont
  {Bairey}}, \bibinfo {author} {\bibfnamefont {C.}~\bibnamefont {Guo}},
  \bibinfo {author} {\bibfnamefont {D.}~\bibnamefont {Poletti}}, \bibinfo
  {author} {\bibfnamefont {N.~H.}\ \bibnamefont {Lindner}},\ and\ \bibinfo
  {author} {\bibfnamefont {I.}~\bibnamefont {Arad}},\ }\bibfield  {title}
  {\bibinfo {title} {Learning the dynamics of open quantum systems from local
  measurements},\ }\href {https://doi.org/10.1088/1367-2630/ab75d1} {\bibfield
  {journal} {\bibinfo  {journal} {New Journal of Physics}\ }\textbf {\bibinfo
  {volume} {22}},\ \bibinfo {pages} {032001} (\bibinfo {year}
  {2020})}\BibitemShut {NoStop}%
\bibitem [{\citenamefont {Fishman}\ \emph
  {et~al.}(2022{\natexlab{a}})\citenamefont {Fishman}, \citenamefont {White},\
  and\ \citenamefont {Stoudenmire}}]{SciPostPhysCodeb.4-r0.3}%
  \BibitemOpen
  \bibfield  {author} {\bibinfo {author} {\bibfnamefont {M.}~\bibnamefont
  {Fishman}}, \bibinfo {author} {\bibfnamefont {S.}~\bibnamefont {White}},\
  and\ \bibinfo {author} {\bibfnamefont {E.~M.}\ \bibnamefont {Stoudenmire}},\
  }\bibfield  {title} {\bibinfo {title} {Codebase release 0.3 for itensor},\
  }\href {https://doi.org/10.21468/SciPostPhysCodeb.4-r0.3} {\bibfield
  {journal} {\bibinfo  {journal} {SciPost Phys. Codebases}\ ,\ \bibinfo {pages}
  {4}} (\bibinfo {year} {2022}{\natexlab{a}})},\ \bibinfo {note} {codebase
  release 0.3}\BibitemShut {NoStop}%
\bibitem [{\citenamefont {Fishman}\ \emph
  {et~al.}(2022{\natexlab{b}})\citenamefont {Fishman}, \citenamefont {White},\
  and\ \citenamefont {Stoudenmire}}]{SciPostPhysCodeb.4}%
  \BibitemOpen
  \bibfield  {author} {\bibinfo {author} {\bibfnamefont {M.}~\bibnamefont
  {Fishman}}, \bibinfo {author} {\bibfnamefont {S.}~\bibnamefont {White}},\
  and\ \bibinfo {author} {\bibfnamefont {E.~M.}\ \bibnamefont {Stoudenmire}},\
  }\bibfield  {title} {\bibinfo {title} {The itensor software library for
  tensor network calculations},\ }\href
  {https://doi.org/10.21468/SciPostPhysCodeb.4} {\bibfield  {journal} {\bibinfo
   {journal} {SciPost Phys. Codebases}\ ,\ \bibinfo {pages} {4}} (\bibinfo
  {year} {2022}{\natexlab{b}})}\BibitemShut {NoStop}%
\bibitem [{\citenamefont {Schollw{\"o}ck}(2011)}]{SCHOLLWOCK201196}%
  \BibitemOpen
  \bibfield  {author} {\bibinfo {author} {\bibfnamefont {U.}~\bibnamefont
  {Schollw{\"o}ck}},\ }\bibfield  {title} {\bibinfo {title} {The density-matrix
  renormalization group in the age of matrix product states},\ }\href
  {https://doi.org/10.1016/j.aop.2010.09.012} {\bibfield  {journal} {\bibinfo
  {journal} {Annals of Physics}\ }\textbf {\bibinfo {volume} {326}},\ \bibinfo
  {pages} {96} (\bibinfo {year} {2011})}\BibitemShut {NoStop}%
\bibitem [{\citenamefont {Davis}\ and\ \citenamefont
  {Kahan}(1970)}]{Davis1970}%
  \BibitemOpen
  \bibfield  {author} {\bibinfo {author} {\bibfnamefont {C.}~\bibnamefont
  {Davis}}\ and\ \bibinfo {author} {\bibfnamefont {W.~M.}\ \bibnamefont
  {Kahan}},\ }\bibfield  {title} {\bibinfo {title} {The rotation of
  eigenvectors by a perturbation. iii},\ }\href
  {https://doi.org/10.1137/0707001} {\bibfield  {journal} {\bibinfo  {journal}
  {SIAM Journal on Numerical Analysis}\ }\textbf {\bibinfo {volume} {7}},\
  \bibinfo {pages} {1} (\bibinfo {year} {1970})}\BibitemShut {NoStop}%
\bibitem [{\citenamefont {Wedin}(1972)}]{Wedin1972}%
  \BibitemOpen
  \bibfield  {author} {\bibinfo {author} {\bibfnamefont {P.-{\AA}.}\
  \bibnamefont {Wedin}},\ }\bibfield  {title} {\bibinfo {title} {Perturbation
  bounds in connection with singular value decomposition},\ }\href
  {https://doi.org/10.1007/BF01932678} {\bibfield  {journal} {\bibinfo
  {journal} {BIT}\ }\textbf {\bibinfo {volume} {12}},\ \bibinfo {pages} {99}
  (\bibinfo {year} {1972})}\BibitemShut {NoStop}%
\bibitem [{\citenamefont {Weyl}(1912)}]{Weyl1912}%
  \BibitemOpen
  \bibfield  {author} {\bibinfo {author} {\bibfnamefont {H.}~\bibnamefont
  {Weyl}},\ }\bibfield  {title} {\bibinfo {title} {Das asymptotische
  verteilungsgesetz der eigenwerte linearer partieller differentialgleichungen
  (mit einer anwendung auf die theorie der hohlraumstrahlung)},\ }\href
  {http://eudml.org/doc/158545} {\bibfield  {journal} {\bibinfo  {journal}
  {Mathematische Annalen}\ }\textbf {\bibinfo {volume} {71}},\ \bibinfo {pages}
  {441} (\bibinfo {year} {1912})}\BibitemShut {NoStop}%
\bibitem [{\citenamefont {Stewart}(1990)}]{Stewart1990}%
  \BibitemOpen
  \bibfield  {author} {\bibinfo {author} {\bibfnamefont {G.~W.}\ \bibnamefont
  {Stewart}},\ }\bibfield  {title} {\bibinfo {title} {Perturbation theory for
  the singular value decomposition},\ }in\ \href@noop {} {\emph {\bibinfo
  {booktitle} {SVD and Signal Processing II: Algorithms, Analysis and
  Applications}}}\ (\bibinfo  {publisher} {Elsevier},\ \bibinfo {address}
  {Amsterdam},\ \bibinfo {year} {1990})\ pp.\ \bibinfo {pages}
  {99--109}\BibitemShut {NoStop}%
\bibitem [{\citenamefont {Chan}\ and\ \citenamefont
  {Van~Voorhis}(2005)}]{Chan2005}%
  \BibitemOpen
  \bibfield  {author} {\bibinfo {author} {\bibfnamefont {G.~K.-L.}\
  \bibnamefont {Chan}}\ and\ \bibinfo {author} {\bibfnamefont {T.}~\bibnamefont
  {Van~Voorhis}},\ }\bibfield  {title} {\bibinfo {title} {Density-matrix
  renormalization-group algorithms with nonorthogonal orbitals and
  non-hermitian operators, and applications to polyenes},\ }\bibfield
  {journal} {\bibinfo  {journal} {The Journal of Chemical Physics}\ }\textbf
  {\bibinfo {volume} {122}},\ \href {https://doi.org/10.1063/1.1899124}
  {10.1063/1.1899124} (\bibinfo {year} {2005})\BibitemShut {NoStop}%
\bibitem [{\citenamefont {Casagrande}\ \emph {et~al.}(2021)\citenamefont
  {Casagrande}, \citenamefont {Poletti},\ and\ \citenamefont
  {Landi}}]{Casagrande_2021}%
  \BibitemOpen
  \bibfield  {author} {\bibinfo {author} {\bibfnamefont {H.~P.}\ \bibnamefont
  {Casagrande}}, \bibinfo {author} {\bibfnamefont {D.}~\bibnamefont
  {Poletti}},\ and\ \bibinfo {author} {\bibfnamefont {G.~T.}\ \bibnamefont
  {Landi}},\ }\bibfield  {title} {\bibinfo {title} {Analysis of a density
  matrix renormalization group approach for transport in open quantum
  systems},\ }\href {https://doi.org/10.1016/j.cpc.2021.108060} {\bibfield
  {journal} {\bibinfo  {journal} {Computer Physics Communications}\ }\textbf
  {\bibinfo {volume} {267}},\ \bibinfo {pages} {108060} (\bibinfo {year}
  {2021})}\BibitemShut {NoStop}%
\bibitem [{\citenamefont {Yamamoto}\ \emph {et~al.}(2022)\citenamefont
  {Yamamoto}, \citenamefont {Nakagawa}, \citenamefont {Tezuka}, \citenamefont
  {Ueda},\ and\ \citenamefont {Kawakami}}]{Yamamoto2022}%
  \BibitemOpen
  \bibfield  {author} {\bibinfo {author} {\bibfnamefont {K.}~\bibnamefont
  {Yamamoto}}, \bibinfo {author} {\bibfnamefont {M.}~\bibnamefont {Nakagawa}},
  \bibinfo {author} {\bibfnamefont {M.}~\bibnamefont {Tezuka}}, \bibinfo
  {author} {\bibfnamefont {M.}~\bibnamefont {Ueda}},\ and\ \bibinfo {author}
  {\bibfnamefont {N.}~\bibnamefont {Kawakami}},\ }\bibfield  {title} {\bibinfo
  {title} {Universal properties of dissipative tomonaga-luttinger liquids: Case
  study of a non-hermitian xxz spin chain},\ }\href
  {https://doi.org/10.1103/PhysRevB.105.205125} {\bibfield  {journal} {\bibinfo
   {journal} {Phys. Rev. B}\ }\textbf {\bibinfo {volume} {105}},\ \bibinfo
  {pages} {205125} (\bibinfo {year} {2022})}\BibitemShut {NoStop}%
\bibitem [{\citenamefont {Zhong}\ \emph {et~al.}(2025)\citenamefont {Zhong},
  \citenamefont {Pan}, \citenamefont {Lin}, \citenamefont {Wang},\ and\
  \citenamefont {Hu}}]{Zhong25}%
  \BibitemOpen
  \bibfield  {author} {\bibinfo {author} {\bibfnamefont {P.}~\bibnamefont
  {Zhong}}, \bibinfo {author} {\bibfnamefont {W.}~\bibnamefont {Pan}}, \bibinfo
  {author} {\bibfnamefont {H.}~\bibnamefont {Lin}}, \bibinfo {author}
  {\bibfnamefont {X.}~\bibnamefont {Wang}},\ and\ \bibinfo {author}
  {\bibfnamefont {S.}~\bibnamefont {Hu}},\ }\bibfield  {title} {\bibinfo
  {title} {Density matrix renormalization group algorithm for non-hermitian
  systems},\ }\href {https://doi.org/10.1103/5vnl-w9p4} {\bibfield  {journal}
  {\bibinfo  {journal} {Phys. Rev. Lett.}\ }\textbf {\bibinfo {volume} {135}},\
  \bibinfo {pages} {106502} (\bibinfo {year} {2025})}\BibitemShut {NoStop}%
\bibitem [{\citenamefont {Van~Damme}\ \emph {et~al.}(2024)\citenamefont
  {Van~Damme}, \citenamefont {Haegeman}, \citenamefont {McCulloch},\ and\
  \citenamefont {Vanderstraeten}}]{VanDamme2024}%
  \BibitemOpen
  \bibfield  {author} {\bibinfo {author} {\bibfnamefont {M.}~\bibnamefont
  {Van~Damme}}, \bibinfo {author} {\bibfnamefont {J.}~\bibnamefont {Haegeman}},
  \bibinfo {author} {\bibfnamefont {I.}~\bibnamefont {McCulloch}},\ and\
  \bibinfo {author} {\bibfnamefont {L.}~\bibnamefont {Vanderstraeten}},\
  }\bibfield  {title} {\bibinfo {title} {Efficient higher-order matrix product
  operators for time evolution},\ }\href
  {https://doi.org/10.21468/SciPostPhys.17.5.135} {\bibfield  {journal}
  {\bibinfo  {journal} {SciPost Phys.}\ }\textbf {\bibinfo {volume} {17}},\
  \bibinfo {pages} {135} (\bibinfo {year} {2024})}\BibitemShut {NoStop}%
\bibitem [{\citenamefont {Dai}\ \emph {et~al.}(2025)\citenamefont {Dai},
  \citenamefont {Wu}, \citenamefont {Wang},\ and\ \citenamefont
  {Zaletel}}]{Dai2025}%
  \BibitemOpen
  \bibfield  {author} {\bibinfo {author} {\bibfnamefont {Z.}~\bibnamefont
  {Dai}}, \bibinfo {author} {\bibfnamefont {Y.}~\bibnamefont {Wu}}, \bibinfo
  {author} {\bibfnamefont {T.}~\bibnamefont {Wang}},\ and\ \bibinfo {author}
  {\bibfnamefont {M.~P.}\ \bibnamefont {Zaletel}},\ }\bibfield  {title}
  {\bibinfo {title} {Fermionic isometric tensor network states in two
  dimensions},\ }\href {https://doi.org/10.1103/PhysRevLett.134.026502}
  {\bibfield  {journal} {\bibinfo  {journal} {Phys. Rev. Lett.}\ }\textbf
  {\bibinfo {volume} {134}},\ \bibinfo {pages} {026502} (\bibinfo {year}
  {2025})}\BibitemShut {NoStop}%
\bibitem [{\citenamefont {Vanthilt}\ \emph {et~al.}(2026)\citenamefont
  {Vanthilt}, \citenamefont {Damme}, \citenamefont {Haegeman}, \citenamefont
  {McCulloch},\ and\ \citenamefont {Vanderstraeten}}]{vanthilt2026}%
  \BibitemOpen
  \bibfield  {author} {\bibinfo {author} {\bibfnamefont {V.}~\bibnamefont
  {Vanthilt}}, \bibinfo {author} {\bibfnamefont {M.~V.}\ \bibnamefont {Damme}},
  \bibinfo {author} {\bibfnamefont {J.}~\bibnamefont {Haegeman}}, \bibinfo
  {author} {\bibfnamefont {I.~P.}\ \bibnamefont {McCulloch}},\ and\ \bibinfo
  {author} {\bibfnamefont {L.}~\bibnamefont {Vanderstraeten}},\ }\href
  {https://arxiv.org/abs/2605.21597} {\bibinfo {title} {Matrix product operator
  encodings of the magnus expansion and dyson series}} (\bibinfo {year}
  {2026}),\ \Eprint {https://arxiv.org/abs/2605.21597} {arXiv:2605.21597
  [quant-ph]} \BibitemShut {NoStop}%
\bibitem [{\citenamefont {Mickiewicz}\ \emph {et~al.}(2026)\citenamefont
  {Mickiewicz}, \citenamefont {Link},\ and\ \citenamefont
  {Strunz}}]{Mickiewicz2026}%
  \BibitemOpen
  \bibfield  {author} {\bibinfo {author} {\bibfnamefont {K.}~\bibnamefont
  {Mickiewicz}}, \bibinfo {author} {\bibfnamefont {V.}~\bibnamefont {Link}},\
  and\ \bibinfo {author} {\bibfnamefont {W.~T.}\ \bibnamefont {Strunz}},\
  }\bibfield  {title} {\bibinfo {title} {Exact floquet dynamics of strongly
  damped driven quantum systems},\ }\href {https://doi.org/10.1103/5z1m-122d}
  {\bibfield  {journal} {\bibinfo  {journal} {Phys. Rev. Lett.}\ }\textbf
  {\bibinfo {volume} {136}},\ \bibinfo {pages} {200201} (\bibinfo {year}
  {2026})}\BibitemShut {NoStop}%
\end{thebibliography}
\end{document}